\documentclass[10pt,conference]{IEEEtran}
\IEEEoverridecommandlockouts
\usepackage[font=footnotesize, skip=2pt]{caption}
\usepackage{enumitem}
\usepackage{tcolorbox}
\tcbuselibrary{theorems}
\newcounter{rq}

\newcounter{arq}

\usepackage{float}
\usepackage{booktabs}
\usepackage{multirow}
\usepackage{cite}
\usepackage{amsmath,amssymb,amsfonts}
\usepackage{algorithmic}
\usepackage{graphicx}
\usepackage{textcomp}
\usepackage[normalem]{ulem}
\usepackage{xcolor}
\usepackage{comment}
\usepackage{algorithmic}
\usepackage{tabularx}
\usepackage{booktabs}
\usepackage[hidelinks]{hyperref}

\def\BibTeX{{\rm B\kern-.05em{\sc i\kern-.025em b}\kern-.08em
    T\kern-.1667em\lower.7ex\hbox{E}\kern-.125emX}}

\newcommand\major[1]{\textcolor{black}{#1}}

\newcommand{\todo}[1]{\textcolor{red}{#1}}

\begin{document}

%\title{Evaluating Ethical Reasoning Agreement in Multi-Modal Generative AI Models for Software Engineering Applications}

\title{Advancing Automated Ethical Profiling in SE:\\a Zero-Shot Evaluation of LLM Reasoning} 
%\title{\major{Evaluating LLM Ethical Reasoning for SE:\\Zero-Shot Evidence for an Ethical Interpreter}}

%Agreement

\author{\IEEEauthorblockN{Patrizio Migliarini}
 \IEEEauthorblockA{\textit{} 
 \textit{University of L'Aquila}\\
 L'Aquila, Italy \\
 patrizio.migliarini@univaq.it}
 \and
 \IEEEauthorblockN{Mashal Afzal Memon}
 \IEEEauthorblockA{\textit{} 
 \textit{University of L'Aquila}\\
 L'Aquila, Italy \\
 mashal.memon@univaq.it}
 \and
 \IEEEauthorblockN{Marco Autili}
 \IEEEauthorblockA{\textit{}
 \textit{University of L'Aquila}\\
 L'Aquila, Italy \\
 marco.autili@univaq.it}
 \and
 \IEEEauthorblockN{Paola Inverardi}
 \IEEEauthorblockA{\textit{Gran Sasso Science Institute} 
 \textit{}\\
 L'Aquila, Italy \\
 paola.inverardi@gssi.it}
% \and
% \and
}
%\author{authors}
\maketitle
\vspace{-1cm}
\begin{abstract}
Large Language Models (LLMs) are increasingly integrated into software engineering (SE) tools for tasks that extend beyond code synthesis, including judgment under uncertainty and reasoning in ethically significant contexts. We present a fully automated framework for assessing ethical reasoning capabilities across 16 LLMs in a zero-shot setting, using 30 real-world ethically charged scenarios. Each model is prompted to identify the most applicable ethical theory to an action, assess its moral acceptability, and explain the reasoning behind their choice. Responses are compared against expert ethicists' choices using inter-model agreement metrics. Our results show that LLMs achieve an average Theory Consistency Rate (TCR) of 73.3\% and Binary Agreement Rate (BAR) on moral acceptability of 86.7\%, with interpretable divergences concentrated in ethically ambiguous cases. A qualitative analysis of free-text explanations reveals strong conceptual convergence across models despite surface-level lexical diversity. These findings support the potential viability of LLMs as ethical inference engines within SE pipelines, enabling scalable, auditable, and adaptive integration of user-aligned ethical reasoning. Our focus is the Ethical Interpreter component of a broader profiling pipeline: we evaluate whether current LLMs exhibit sufficient interpretive stability and theory-consistent reasoning to support automated profiling.
\end{abstract}


\section{Introduction}
\label{sec:intro}

%\noindent\textbf{\major{Autonomous systems and the ethical design challenge.}}
Autonomous systems are increasingly becoming an integral part of our daily lives across diverse domains~\cite{suri2023software,jedlickova2024ensuring}. These systems can operate independently without any human intervention and make decisions acting on behalf of their users~\cite{InsightsISSRE,waldman2019power,sharma2022recent,anderson2018artificial}. Their rapid growth brings both opportunities and challenges. From a software engineering perspective, as these systems become pervasive, a key challenge is designing systems that, beyond meeting technical requirements, also account for ethical considerations~\cite{autili2025engineering,de2024engineering,hendrycks2020aligning,alidoosti2022incorporating,svegliato2021ethically}.

\smallskip
\noindent\textbf{\major{SE ethics.}}
%: principles and alignment mechanisms.}}
Recently, various studies have focused on the ethical implications of these software-intensive systems on individuals and society~\cite{alidoosti2022incorporating,tolmeijer2020implementations,inverardi2019ethics,inverardi2022ethical,cervantes2020toward}. Software engineering ethics encompasses principles and rules that guide engineers' decisions throughout the design and development process~\cite{alidoosti2025exploring}. Various approaches have also been introduced that ensure that systems align with broad ethical values like fairness, transparency, and safety~\cite{jedlivckova2024ethical,bremner2019proactive,winfield2014towards,Winfield2019,alidoosti2021ethics,townsend2022pluralistic}.

\smallskip
%\noindent\textbf{\major{From abstract norms to user-aligned ethical profiles.}}
\noindent\textbf{\major{Ethics operationalization.}}
Moreover, beyond abstract ethical norms, recent studies focus on operationalizing end users' ethical preferences directly into the software engineering process~\cite{winter2019advancing,memon2023automated,MemonAFSI24,donati2024representing,de2024engineering,barn2016you}. Through profiling techniques \major{(}including questionnaires, surveys, product reviews\major{,} etc.\major{)}, these studies propose approaches to capture user ethical preferences, generate their ethical profiles, and integrate \major{them} into these systems~\cite{alfieri2022exosoul,autili2025engineering,cacmInverardi2019,InverardiMP23,DiRuscioIMN24}. The ethical profile is a structured representation of the ethical preferences of the user that autonomous systems can leverage to adjust their behavior and make decisions aligned with the user\major{'s} ethical values~\cite{alfieri2022exosoul}. Embedding user ethical profiles into system design would not only enhance trust and accountability, as highlighted by regulatory bodies including GDPR~\cite{voigt2017eu}, the AI Act~\cite{EU_AI_Act_Proposal}, and the Ethics Guidelines for Trustworthy AI~\cite{ai2019high}, but also ensure that systems reflect and respect the ethical preferences of their users. However, relying on manual input from users to generate their profiles limits their scope and adaptability, as users' ethical preferences vary with the change in context. Hence, it becomes impractical to expect users to provide input for every possible situation, introducing the challenge of automating the generation of ethical profiles.

\smallskip
%\noindent\textbf{\major{LLMs as candidates for ethical reasoning.}}
\noindent\textbf{\major{LLM-based ethical reasoning.}}
Recent advances in generative AI, especially large language models (LLMs), have positioned these models as powerful tools capable of engaging in ethical reasoning~\cite{han2022aligning}. Various studies have explored the use of LLMs to assess their moral reasoning abilities in specific applications~\cite{alshami2023harnessing,DBLP:conf/emnlp/RaoKTAC23,DBLP:conf/iclr/TennantHM25}. Building on this, we take a step toward evaluating whether large language models (LLMs) can effectively reason about ethically significant content in real-life scenarios. To this end, we propose a lightweight, fully automated framework that examines the potential of LLMs to identify ethically relevant information, to support the automated generation of user-aligned ethical profiles, and \major{to integrate them} into software engineering (SE) pipelines.

\smallskip
\noindent\textbf{\major{Setup and RQs.}}
We present 16 LLMs (as shown in Table~\ref{tab:llms_summary}) with 30 ethically charged statements. For each statement, the models are prompted to identify the most applicable ethical theory according to the action detailed in the statement, assess whether the action is morally acceptable according to the selected theory, and explain the reasoning behind their choice. Importantly, as discussed in Section \ref{sec:scenario}, selecting an ethical theory does not imply that the action described in the statement is justified by that theory; rather, it serves as a normative lens through which the moral acceptability of the action is evaluated. This distinction enables meaningful binary judgments (acceptable/unacceptable) based on whether the action complies with the principles of the selected theory. To establish a \major{comparative baseline}, we replicate the same process with three professors, experts with extensive knowledge in applied ethics and philosophy. We then evaluated the responses of LLMs and experts, exploring both alignment and divergence in their judgments and their explanations. To assess the potential of LLMs as ethical reasoning modules in software engineering, we pose the following research questions:

\smallskip
\noindent\textbf{RQ1:} Do LLMs demonstrate the capacity for ethical reasoning when presented with ethically charged scenarios?\\
\textbf{RQ2:} To what extent do different LLMs agree on ethical theories and moral acceptability of the scenarios?\\
\textbf{RQ3:} How do the agreements among LLMs compare to those among the human experts?\\
\textbf{RQ4:} What qualitative characteristics emerge in the explanations produced by LLMs?

\smallskip
\noindent\textbf{\major{Methodology.}}
These questions are designed to evaluate whether current generative models can function not only as isolated agents but also as components in robust, transparent, and auditable decision pipelines for software engineering. To address RQ1, we prompted 16 LLMs with 30 ethical scenarios covering a range of common, real-life contexts that involve ethically charged situations. We then analyzed LLMs' responses to determine their ability to recognize the action described in the scenario, its correspondence to one of the ethical theories (utilitarianism, deontology, and virtue ethics), and determine whether the action described is morally acceptable according to the selected ethical theory. To address RQ2, we computed inter-model agreement using Theory Consistency Rate (TCR) to identify the percentage of prompts for which the models selected the same ethical theory and Binary Agreement Rate \major{metrics} (BAR) to identify the percentage of prompts for which different models agreed on whether the action is morally acceptable in accordance with the selected ethical theory. These metrics assess whether different LLM models apply comparable ethical reasoning structures under identical conditions. To address RQ3, we collected questionnaires from the three professors, expert ethicists, and we presented them with the same scenarios previously shown to the LLMs. We surveyed them by replicating the same process we followed with LLMs. We then compared the experts’ judgments with the responses generated by the LLMs using z-score\major{s}. To address RQ4, we conducted a multi-layered qualitative analysis of the free-text explanations provided by the LLMs. We applied lexical similarity metrics (TF-IDF and cosine similarity), dimensionality reduction (PCA, t-SNE), and topic modeling (LDA) to examine variation in linguistic form and underlying conceptual structure. A manual alignment study further assessed whether the explanations were consistent with the ethical theories selected by the models. This combined approach enabled the evaluation of both the coherence and the diversity of the explanations.

\smallskip
\noindent\textbf{\major{LLM-aided ethical reasoning.}}
From a software engineering perspective, the results indicate that LLMs are capable of serving as modular evaluators of ethical context. They show capabilities that can be utilized as a possible way to automate the generation of user ethical profiles. \major{This paper does not present a full automated ethical profile generator. Instead, it evaluates whether state-of-the-art LLMs can serve as ethical reasoning modules, that is, whether they can consistently select an interpretive moral lens and produce theory-aligned acceptability explanations under zero-shot conditions. Within SE pipelines, this component enables: (i) decision auditing with concise theory-grounded rationales; (ii) triage and escalation when model disagreement signals moral ambiguity; (iii) seed signals for user-aligned profiling, where interpretive patterns accumulate into dynamic profiles.}

\smallskip
\noindent\textbf{\major{Contributions.}}
The main contributions of the paper are:
\begin{itemize}[left=0pt]
\item An automated framework for quantifying agreement and divergence among LLMs on non-trivial reasoning tasks, using ethical scenarios as a representative benchmark.
\item An empirical analysis of the consistency and diversity of 16 LLMs, measuring both classification agreement and the qualitative variety in explanations.
\item An evidence-based discussion on the trade-offs of single-LLM versus multi-LLM approaches for judgment and reasoning in SE support systems.
\item A comparison of the LLM outcomes and those of expert human judgments, identifying areas of alignment and persistent divergence.
\item A fully reproducible experimental pipeline and dataset, with all artifacts released for independent verification.
\end{itemize}

\smallskip
\noindent\textbf{\major{Paper roadmap.}}
The rest of the paper is organized as follows.
Section~\ref{sec:settingcontext} introduces the theoretical foundations of ethical reasoning and motivates the need for automated ethical profiling in SE.
Section~\ref{sec:approach} provides an overview of our evaluation framework and its main components.
Section~\ref{sec:quantitative} reports inter-model and human–LLM agreement metrics to assess consistency in ethical reasoning.
Section~\ref{sec:qualitative} analyzes the LLM-generated explanations through lexical, conceptual, and theory-alignment perspectives.
Section~\ref{sec:discuss} discusses the implications of our findings, identifies limitations, and outlines practical use cases.
Section~\ref{sec:related} reviews related work on ethical AI, moral reasoning, and LLM evaluation.
Finally, Section~\ref{sec:conclude} concludes the paper and highlights directions for future research.

\section{Setting the Context}
\label{sec:settingcontext}

In this section, we introduce background concepts that form the basis of our approach and frame this work inside the umbrella project Exosoul~\cite{AutiliACCESS19} which is about protecting citizens' ethics and privacy in the digital world.

Understanding how individuals make ethical decisions in real-life scenarios is a crucial step in designing systems that can adapt to their users' ethical preferences~\cite{autili2025engineering}.
While various approaches have proposed the use of LLMs for moral reasoning, they are primarily designed to fine-tune the models and test their reasoning capabilities rather than assessing their inherent capacity for moral reasoning~\cite{han2022aligning}~\cite{ICSE2025PAPER}. Our end goal, however, is to automatically generate user ethical profiles utilizing LLMs, which reflect how individual users would act in an ethically charged situation. Hence, this involves evaluating LLMs' ethical reasoning capabilities.

\textbf{The choice of questionnaire.} To develop our approach, we build on top of the questionnaire introduced in~\cite{alfieri2022exosoul} and revised in~\cite{alfieri2023ethical}. The questionnaire is introduced by a multidisciplinary group, including ethicists, philosophers, and researchers engaged in applied ethics and cognitive science, to collect ethical preferences from users in real-life situations. The questionnaire is composed of questions that reflect everyday moral dilemmas, designed to manually generate users' ethical profiles from their responses. To adapt this questionnaire for our approach, we generated declarative statements from these questions (further discussed in Section~\ref{sec:approach}). This translation allows us to leverage the questionnaire in a more structured format, making it suitable for evaluating whether LLMs can identify the presence of ethically charged actions within these statements. The original questionnaire and all the used statements are presented in the replication package.\footnote{\label{replication}{https://github.com/ASE25authors/ase-aep}}

%\footnote{\label{replication}\todo{The replication package can be found at this link: https://bit.ly/3HgiquG}}

\textbf{The choice of theories.} Ethical considerations in the fields of computer science and software engineering have become increasingly important as technology advances~\cite{singh2023approaches}. The extensive use of artificial intelligence and machine learning has further made it essential to evaluate the ethical implications of such technologies~\cite{UNESCOGuidelines,ryan2020artificial,anderson2018artificial}. Various ethical theories can be employed to ensure their ethical development and application~\cite{singh2023approaches,tolmeijer2020implementations}. In this work, we evaluate whether LLMs possess ethical reasoning capabilities and identify the ethical significance of actions by evaluating their responses against three foundational moral theories: utilitarianism, virtue ethics, and deontology. The selection of these theories is based not only on their significant influence but also on their widespread adoption across both applied ethics and the broader AI and machine ethics literature~\cite{tolmeijer2020implementations,guarini2013introduction,anderson2020machine,o2012review,jedlivckova2024ethical}. Among the selected theories, Utilitarianism evaluates actions based on their outcomes, aiming to maximize overall well-being or happiness~\cite{mill2016utilitarianism}. Virtue ethics focuses on the moral character and intentions~\cite{hursthouse2017virtue}. Deontology judges actions based on adherence to moral duties or principles, regardless of the outcome~\cite{sep-ethics-deontological}. The statements we used in our approach are grounded in real-life situations representing an ethical dilemma, %each of which can be interpreted through one or more of these ethical theories, making these theories well-suited for the scope of our work.
in which each decision taken may correspond to one or more of these ethical theories, highlighting the relevance of these ethical theories to our work. %answers Rev C on why only one “best apply” and prevents ambiguity already in Sec. II
\major{In our setting, multiple theories may be applicable to the same scenario. We therefore treat the selected theory as a normative lens, an interpretive perspective used to frame the subsequent yes/no acceptability judgment. The task is not to assert the “true” or exclusive theory, but to elicit structured moral reasoning under an explicitly plural and interpretive design.}

%\textbf{Ethical profiles.} An ethical profile is a set of a user's ethical preferences. Ethical profiles are dynamic and context-sensitive, enabling the generation of context-specific profiles representing how a user would act in a specific context. Several studies have proposed methods for generating users’ ethical profiles using tools such as questionnaires, surveys, product reviews, and social media interactions~\cite{boja2019user,gilbert2023rise,dong2021profiling,alfieri2022exosoul,alfieri2023ethical}. This is precisely where project [anonymized] comes into play with one of its main objectives: automating the generation of ethical profile. Figure 1 briefly overviews the overall approach of the project for what concerns the ethical profile creation... 
%\pat{describe deeply}
%pipeline showing the main components to translate multimedia input from the user to 

\textbf{Ethical profiling.} An ethical profile is a structured representation of a user's ethical preferences, designed to reflect how the user would respond to ethically significant decisions in context-specific scenarios. These profiles are inherently dynamic and situated, evolving with the user’s behavior and values. Prior studies have proposed their construction through explicit elicitation, such as surveys, questionnaires, or review analysis~\cite{boja2019user,gilbert2023rise,dong2021profiling,alfieri2022exosoul,alfieri2023ethical}. %yet these methods are not designed to scale or adapt to real-time decision-making.
However, the continuous and manual input required from the user limits their scope and adaptability, highlighting the need to automate the generation of ethical profiles.

\begin{figure}[ht]
%\hspace*{-1.2em}
\includegraphics[width=0.49\textwidth]{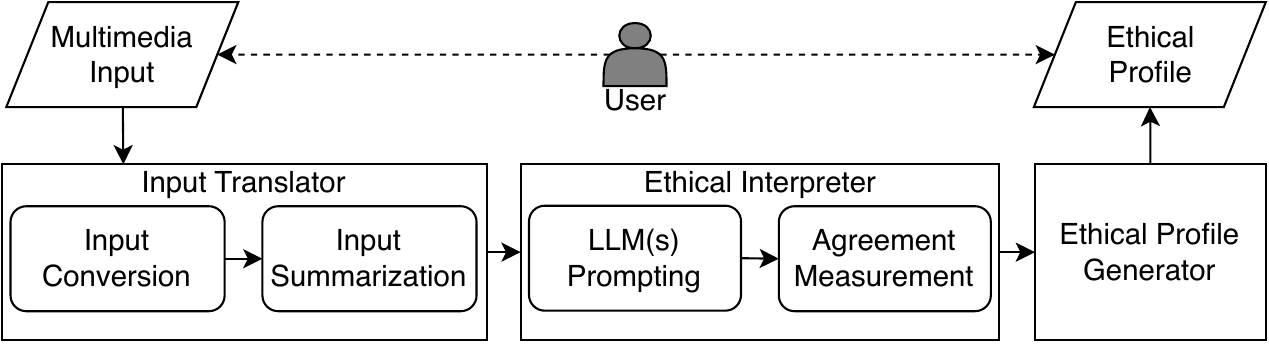}
\caption{Automated Ethical Profile Generation}
\label{fig:approach1}
\end{figure}

The broader vision of project Exosoul~\cite{AutiliACCESS19} is to propose a modular approach to protect and empower individuals in asymmetric interactions with complex digital environments. The primary goal of the project is to mediate ethical, privacy-related, and social frictions through adaptive components that infer, negotiate, and operationalize the user's ethical stance in context. Figure~\ref{fig:approach1} shows one technical subcomponent of the approach: a pipeline for the automated interpretation of ethically relevant content and generation of the user's ethical profile. The pipeline comprises seven modules. \textit{Multimedia Input} collects inputs from user activity or environment, such as video, audio, and text. \textit{Input Translator} performs
\textit{Input Conversion} to convert the inputs into symbolic representations and \textit{Input Summarization} to condense relevant content into structured prompts. \textit{Ethical Interpreter} classifies the summarized input by theory, acceptability, and explanation using \textit{LLM(s) Prompting} to provide general-purpose interpretive ethical outputs, and \textit{Agreement Measurement} compares multiple reasoning traces to assess consistency. \textit{Ethical Profile Generator} aggregates these outputs to synthesize the user’s ethical profile. 

This paper concerns specifically the \textit{Ethical Interpreter} component and aims to evaluate whether current LLMs exhibit the reasoning capability, internal coherence, and interpretive stability required to support this component. Thus, the focus of this paper is not on ethical profile generation, rather the focus is on empirically %validate
\major{validating} the core interpretive layer necessary for supporting the automation of the ethical profile generation.

%However, these approaches often produce either generic profiles that lack contextual depth or context-specific profiles that are impractical, as they require users to provide input for each situation. Therefore, the end goal of our approach is to explore the ethical reasoning capabilities of LLMs and use them to automatically generate ethical profiles by analyzing their behavior.

%\vspace{0.2cm}
%\subsubsection{Related Work} 

%In this paper, we focus on the ethics interpreter component and explore the ethical reasoning capabilities of LLMs and use them to automatically generate ethical profiles by analyzing their behavior.
\section{Approach Overview}
\label{sec:approach}

\begin{figure}[ht]
\includegraphics[width=0.49\textwidth]{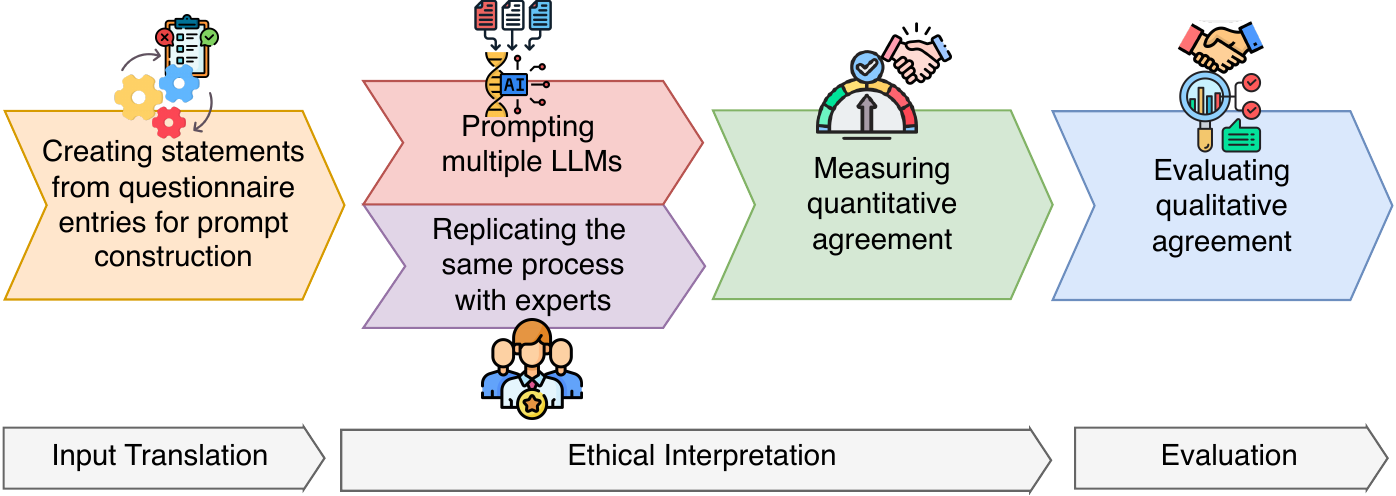}
%\caption{Approach Pipeline}
\caption{Approach overview of this research}
\label{fig:approach2}
\end{figure}

We propose a fully automated, experimental pipeline to compare the ability of LLMs to reason over ethically charged scenarios in zero-shot settings, to support the automated generation of user ethical profiles in software engineering applications. Figure~\ref{fig:approach2} illustrates the overview of the approach.

\textit{Creating statements from questionnaire entries.} The process begins by adapting 30 questions -- drawn from and inspired by the ethical profiling questionnaires used in the study~\cite{alfieri2023ethical} -- into concise scenario statements, each presenting a specific moral dilemma and reflecting a choice that has already been made.

\major{We provide three questionnaire items that were administered to both the LLMs and the human experts:}
\begin{itemize}[left=0pt]
    \item \major{A person helps a postal clerk manually assign queue numbers during a system failure.}    \item \major{A user donates a small amount to Wikipedia after reading a request for funding.}
    \item \major{Someone finds a wallet with €1,000 and no ID and turns it in to the police.}
\end{itemize}

\major{Full corpus of items, prompts, raw responses, and expert annotations is released in the replication package.}

\textit{Prompting multiple LLMs.} Each scenario is presented as a consistently templated prompt shown in Section~\ref{sec:prompting} to 16 state-of-the-art language models, spanning both proprietary and open-source families, using deterministic inference settings as specified in the LLMs documentation.

\textit{Replicating the same process with experts.} Three expert ethicists independently respond to the same set of scenario prompts, enabling direct comparison between model and human ethical judgments.

\textit{Measuring quantitative agreement.} Model and expert responses are analyzed using Theory Consistency Rate (TCR) and Binary Agreement Rate (BAR), alongside standardized z-score normalization, to assess inter-model and human-model consistency.

\begin{table}[H]
  \centering
  \footnotesize
  \setlength{\tabcolsep}{4pt}
  \caption{LLM convergence for the three illustrative items.}
  \label{tab:llm_examples_tcr_bar}
  \begin{tabular}{lcc}
    \toprule
    Item & TCR & BAR \\
    \midrule
    1 & 43.75\% (Utilitarianism) & 100.00\% (YES) \\
    2 & 50.00\% (Utilitarianism) & 93.75\% (YES) \\
    3 & 75.00\% (Deontology)     & 100.00\% (YES) \\
    \bottomrule
  \end{tabular}
\end{table}

\begin{table}[H]
  \centering
  \footnotesize
  \setlength{\tabcolsep}{4pt}
  \caption{Experts convergence for the three illustrative items.}
  \label{tab:expert_examples_tcr_bar}
  \begin{tabular}{lcc}
    \toprule
    Item & TCR & BAR \\
    \midrule
    1 & 66.67\% (Virtue ethics) & 100.00\% (YES) \\
    2 & 100.00\% (Utilitarianism) & 66.67\% (YES) \\
    3 & 66.67\% (Deontology) & 66.67\% (YES) \\
    \bottomrule
  \end{tabular}
\end{table}
\major{In Table~\ref{tab:llm_examples_tcr_bar} and Table~\ref{tab:expert_examples_tcr_bar} we provide the TCR and BAR results for the three questionnaire items provided above. The full set of results is released in the replication package.}

\textit{Evaluating qualitative agreement.} The provided explanations undergo multiple qualitative %analysis
\major{analyses}, including TF-IDF-based lexical similarity, topic modeling, and clustering, to assess reasoning consistency and conceptual alignment.
\major{For each of the three items above, we provide brief explanations from both the LLMs and the human experts:}
\begin{itemize}[left=0pt]
    \item \major{LLM: Assists in resolving system failure, promotes efficiency}
    \item \major{LLM: Shows generosity and support for a valuable resource}
    \item \major{LLM: Acts with respect for others' property and integrity}
    \item \major{Expert: virtue ethics: if the act was in the nature of the individual}
    \item \major{Expert: you pay a little you get a lot. all theories could apply with different reasons}
    \item \major{Expert: no one would judge you, so it is just the way you are deontology}
\end{itemize}

\major{Full explanations set is released in the replication package.}

The \textit{Input Translation, Ethical Interpretation, and Evaluation} pipeline as a whole supports the translation of raw questionnaire data into structured prompts, automated ethical reasoning and explanation generation, and multi-level evaluation of interpretive quality and agreement across LLMs, following the pattern of the modules shown in Figure~\ref{fig:approach1}.

\begin{comment}

\noindent\textit{Ethical Questionnaire.} A set of 30 ethically charged scenarios derived from the Exosoul questionnaire~\cite{alfieri2022exosoul}, designed to reflect recurring moral dilemmas in SE contexts such as responsibility, transparency, and fairness.
\textit{Input (Statements).} Each scenario is formatted into a prompt with three required outputs: (i) selection of the most applicable ethical theory (utilitarianism, deontology, or virtue ethics), (ii) binary judgment on whether the action is morally acceptable, and (iii) free-text explanation.
\textit{Multiple LLMs.} Sixteen leading LLMs, spanning proprietary and open-source models, are evaluated using a shared prompt template and deterministic generation settings.
\textit{Expert Ethicists.} For reference, three expert ethicists independently annotated each scenario using the same protocol, enabling direct comparison with model responses. %\mashal{above we say 5 experts everywhere, so 3 or 5?} \pat{we have 3 questionnaires from experts, so 3.}
\textit{Quantitative Agreement Measurement.} Model outputs are analyzed using %both 
  quantitative agreement metrics (TCR, BAR, z-scores)
\textit{Qualitative Agreement Evaluation}: Qualitative linguistic techniques (TF-IDF similarity, topic modeling, clustering) to assess reasoning consistency and explanation quality.
\end{comment}

%\pat{working on it}

All artifacts, including prompts, responses, and evaluation scripts, are available in the replication package\footref{replication} to support reproducibility and benchmarking.

\begin{comment}

\subsection{Ethical Profile Inference}

The modular structure of the pipeline supports downstream integration with ethical user profiling. By aggregating LLM judgments across diverse scenarios, the system can infer a candidate ethical profile representing the user’s alignment with specific theories (e.g., consistent deontological preferences) or value-based patterns (e.g., high acceptance of privacy-preserving actions). This structure enables practical deployment in SE systems such as recommender agents, personal assistants, or autonomous interface controllers. The same approach can also serve as a frontend to audit existing systems for ethical alignment or profile consistency.

\end{comment}

\subsection{Scenario Dataset and Prompt Construction}
\label{sec:scenario}

The evaluation relies on a benchmark of 30 ethically relevant scenarios, adapted from an established ethical profiling questionnaire previously used in applied ethics and cognitive science research~\cite{alfieri2022exosoul,alfieri2023ethical}. These scenarios were selected for their ability to reflect already taken decisions that are rooted in general moral dilemmas.
%, also represent recurring challenges in software engineering contexts, such as responsibility allocation, privacy management, compliance with policy, and fairness in user interactions. 
% 
Each scenario was expressed as a declarative statement describing a concrete action or choice, carefully phrased to minimize ambiguity while retaining enough realism to engage different dimensions of ethical reasoning. The process involved translating original survey items into concise prompts and validating their clarity through expert review.
%Our evaluation is grounded in a benchmark of 30 ethically charged scenarios adapted from a validated questionnaire in applied ethics and cognitive science~\cite{alfieri2022exosoul}. While originally domain-agnostic, these scenarios were selected for their capacity to highlight ethical dimensions that frequently arise (directly or analogically) within software engineering practice. Specifically, they address themes such as data privacy, responsible system use, fairness in interaction design, accountability in collaborative settings, and ethical behavior in digital work environments. Each scenario is phrased as a concise declarative statement describing a specific action or decision, designed to elicit structured ethical judgment. This structure allows us to systematically examine whether and how LLMs exhibit ethical sensitivity across a spectrum of situations that mirror the ethical complexity encountered in real-world SE contexts.
Examples of %possible adaptations of 
the scenarios used in the study include: 
\begin{itemize}[left=0pt]
\begin{comment}
    \item\textit{A developer refuses to use unlicensed software for an urgent deadline.} 
    \item\textit{A project manager discloses a minor security vulnerability to all users, despite internal advice to keep it confidential.} 
    \item\textit{An employee deletes private information found on a shared drive instead of reporting it.}
\end{comment}

\item A freelancer downloads expensive software illegally to complete an urgent work project.\\
\textit{Reflects real-world dilemmas regarding license compliance, project pressure, and ethical tool use.}

\item A traveler accepts all cookies when buying a flight online.\\
\textit{Raises issues of privacy consent, dark patterns, and ethical user experience design in software systems.}

\item A customer finds a USB stick in a café and plugs it into their own laptop out of curiosity.\\
\textit{Illustrates security risks and responsible behavior in handling external devices.}
\end{itemize}

% clarifies the distinction already discussed in the text, now made front-and-center

\major{Selecting a theory does not imply that the action is justified by that theory; rather, it specifies the lens relative to which the binary acceptability is evaluated. For instance, models may split between deontology and utilitarianism as the best lens for a piracy scenario, yet still converge that the action is not acceptable given either lens. This distinction is central to our evaluation design.}

\smallskip

\begin{comment}

As detailed in subsection \ref{sec:prompting}, for each scenario both LLMs and human experts received the same prompt, structured in three parts consisting in identification of the most appropriate ethical theory (utilitarianism, deontology, or virtue ethics), binary judgment on whether the action is morally acceptable, and a brief free-text explanation for the answer. This format supports structured comparison across models and experts, and enables quantitative as well as qualitative analysis of both outcome and explanation. The complete list of scenarios is provided in the replication package.
\end{comment}

As detailed in subsection~\ref{sec:prompting}, each scenario was presented to both LLMs and human experts using a shared prompt, structured in three parts: (i) selection of the ethical theory that best frames the action (utilitarianism, deontology, or virtue ethics), (ii) a binary judgment on whether the action is morally acceptable under that theory, and (iii) a concise explanation. Crucially, the ethical theory is treated not as a justification mechanism but as a normative lens: an interpretive perspective that informs how moral acceptability is evaluated. The binary response (YES/NO) is therefore relative to the internal logic of the selected theory, that is, whether the action adheres to or violates the principles it prescribes. For instance, the same action may be deemed unacceptable from a deontological standpoint (due to duty violation), yet potentially acceptable from a utilitarian perspective (if it maximizes positive outcomes). To illustrate this point, consider the scenario: \textit{“A freelancer downloads expensive software illegally to complete an urgent work project.”} Despite a relatively low Theory Consistency Rate (TCR) of 37.5\% with LLMs split between deontology and utilitarianism 93.75\% of models judged the action to be morally unacceptable. One LLM selected deontology and explained its judgment as follows: \textit{“Disregards others' intellectual property rights and moral principles.”} This case shows that identifying a theory does not entail endorsement of the action; rather, the binary judgment reflects whether the action conforms to the moral rules of that framework. As further discussed in Section~\ref{sec:qualitative}, we observed occasional misalignments between theory selection, binary judgment, and free-text explanation. These inconsistencies were explicitly analyzed to assess the internal coherence of LLM ethical reasoning.

\subsection{LLM Pool and Prompting Protocol}
\label{sec:prompting}

The evaluation included 16 large language models covering a broad spectrum of architectural families, release dates, and access modalities. This diversity was intended to reflect both widely deployed commercial models and state-of-the-art open-source systems available to practitioners and researchers. The pool comprises both API-accessible and locally executable models, as summarized in Table~\ref{tab:llms_summary}.

\begin{table}[h]
\centering
\caption{Overview of Evaluated LLMs}
\label{tab:llms_summary}
\begin{tabular}{llll}
\toprule
Model & Provider & Access & Parameters \\
\midrule
GPT-4o          & OpenAI      & API      & N/A \\
Claude 3.7 Sonnet   & Anthropic    & API      & N/A \\
Gemini 2.5 Pro      & Google       & API      & N/A \\
Command R+          & Cohere       & API      & N/A \\
Mistral             & Mistral AI   & API      & 7B/8x22B (MoE) \\
Grok 3              & xAI          & API      & N/A \\
Qwen 3              & Alibaba      & API      & 235B \\
LLaMA 2 Chat 7B     & Meta         & API    & 7B \\
LLaMA 3 8B          & Meta         & Local    & 8B \\
LLaMA 3.2 3B        & Meta         & Local    & 3B \\
DeepSeek-R1 Distill & DeepSeek     & Local    & 8B \\
Hermes (OpenHermes) & Nous         & Local    & 7B \\
Orca 2 Full         & Microsoft    & Local    & 13B \\
Reasoner V1         & Community    & Local    & 7B \\
Ghost 7B v0.9.1     & Community    & Local    & 7B \\
Phi-3 Mini Instruct & Microsoft    & Local    & 3.8B \\
\bottomrule
\end{tabular}
\end{table}

Each model was prompted with the same set of 30 scenarios, using an identical input template and zero-shot configuration to ensure comparability. For all API-accessible models, the default deterministic setting was used (temperature set to 0.2 unless otherwise specified by the provider). Local models were run with their recommended default parameters, and all outputs were collected automatically to avoid human bias. The prompt structure presented to each model is:

\smallskip

\begin{quote}
\scriptsize\ttfamily
Given the following scenario: [SCENARIO TEXT] \\
1) Which ethical theory best applies to this situation: utilitarianism, deontology, or virtue ethics?\\
2) Based on the theory you selected, is the action morally acceptable (yes/no)?\\
3) Provide a brief explanation.
\normalfont
\end{quote}

\smallskip

No additional system or context instructions were provided, and no prior examples were shown to the models, reflecting a strict zero-shot approach. All responses were collected in plain text and processed using a multi-script analysis pipeline.

%responds to Rev B Q1 and Rev A on prompt sensitivity

\major{Design rationale (zero-shot): we adopt a strict zero-shot setting to probe intrinsic ethical reasoning capacity under a single, shared prompt. Alternative strategies (one-/few-shot, chain-of-thought, formatting variants) are known to modulate outputs; we deliberately defer such prompt-sensitivity studies to future work to avoid conflating capacity with prompt engineering.}

\major{Determinism and parameters: for API models, we set temperature to 0.2 (or provider defaults when lower/hard-coded) to reduce sampling variance while preserving non-trivial reasoning; local models follow recommended deterministic settings. This balances repeatability and expressivity and makes inter-model comparisons fairer under identical inputs.}
%\pat{added}
\section{Quantitative Analysis}
\label{sec:quantitative}

%\subsection{Experimental Protocol and Metrics}

%\textbf{Experimental Protocol and Metrics.} 
For each scenario in the dataset, all 16 LLMs were queried independently using the shared prompt structure. Each response was parsed to extract the selected ethical theory, the binary acceptability judgment, and the free-text explanation. The same protocol was applied to a group of 3 human experts in ethics and applied philosophy to provide a comparative baseline. Agreement among models was quantified using two primary metrics:

\textit{Theory Consistency Rate (TCR).} \major{The share of models that select the modal theory for a scenario; it measures convergence of interpretive framing. TCR is introduced in this work as a deliberately simple, transparent modal-share indicator; it is not a gold-standard agreement coefficient.}

\textit{Binary Agreement Rate (BAR).} \major{The share of models that agree on the yes/no acceptability given their selected lens; it measures outcome-level consensus.}

\major{We report z-scores for both metrics and visualize thresholds to identify high-variance (ambiguous) scenarios.}

% \begin{itemize}
%     \item \textit{Theory Consistency Rate (TCR):} the percentage of models selecting the modal ethical theory for a given scenario, measuring convergence on moral framing.
%     \item \textit{Binary Agreement Rate (BAR):} the percentage of models agreeing on the binary moral acceptability (yes/no), capturing outcome-level consensus.
% \end{itemize}
To account for scale differences, both metrics are standardized via \textit{z-score} normalization ($z_{\mathrm{TCR}}$ and $z_{\mathrm{BAR}}$), where $\mu$ and $\sigma$ represent the mean and standard deviation of the respective metric across all scenarios:
\smallskip
\[
z_\text{TCR} = \frac{\text{TCR} - \mu_\text{TCR}}{\sigma_\text{TCR}}, \qquad
z_\text{BAR} = \frac{\text{BAR} - \mu_\text{BAR}}{\sigma_\text{BAR}}
\]

A combined agreement score per scenario is computed as:

\[
\text{Combined z-score} = \frac{z_\text{TCR} + z_\text{BAR}}{2}
\]

This provides a unified measure of inter-model consistency across both theoretical and acceptability dimensions. The same procedure was applied to the human experts. All parsing and scoring scripts are included in the replication package.
% required by metareview
\smallskip

\begin{figure}[ht]
\centering
\includegraphics[width=0.49\textwidth]{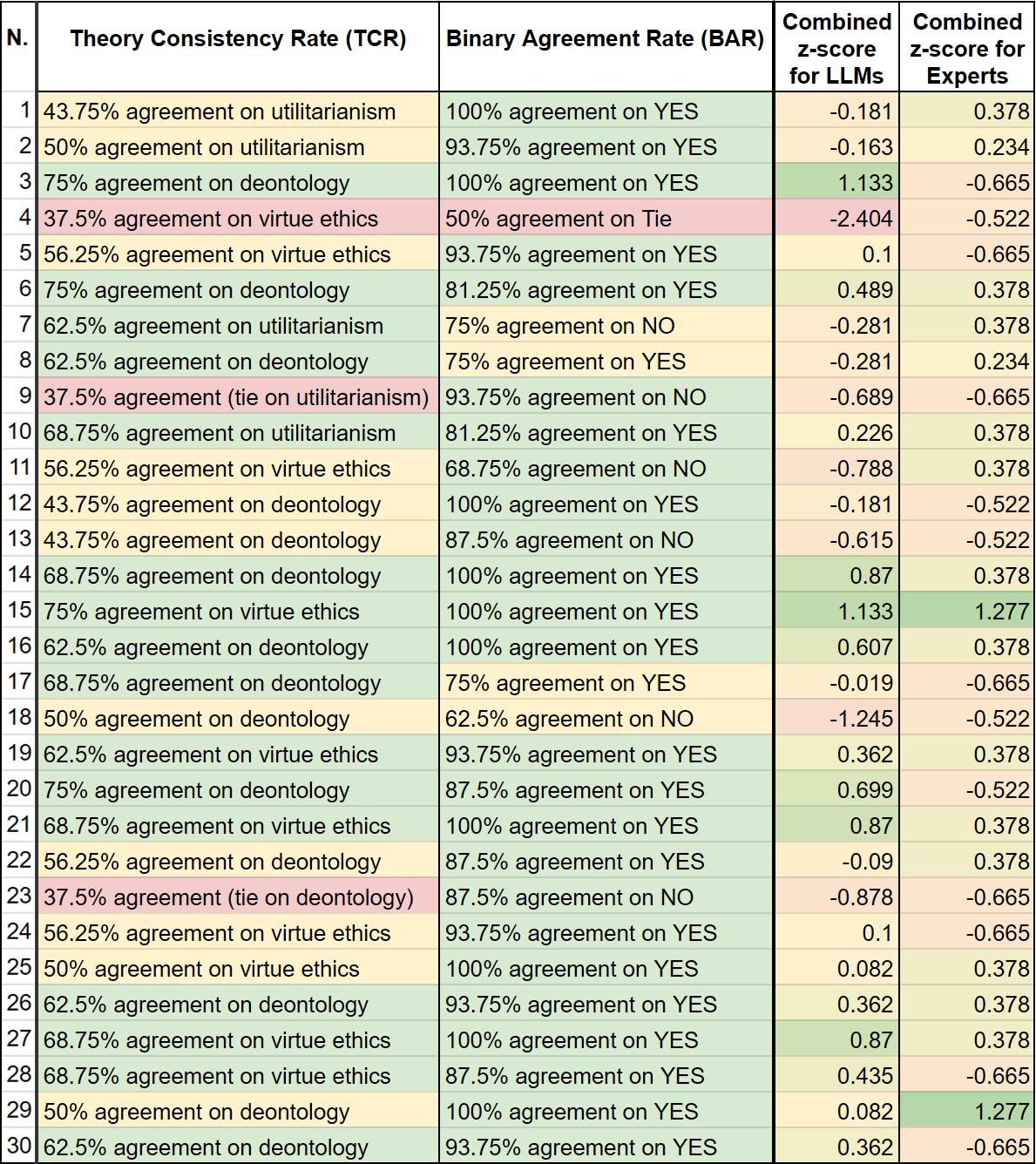}
\caption{LLMs TCR and BAR results with Fleiss' Kappa agreement coloring and Z-Scores with threshold coloring.}
\label{fig:sTCR_BAR_table}
\end{figure}

\textbf{Ethical reasoning capacity (RQ1).}
LLMs %shows
\major{show} ethical reasoning capacities over %ethical-relevant
\major{ethically relevant} scenarios. With a 73.3\% average agreement on ethical theory and 86.7\% on acceptability, models show consistent normative interpretation under zero-shot conditions. Results shown in Figure~\ref{fig:sTCR_BAR_table} suggest that LLMs can produce structured, theory-informed outputs for moral scenarios, despite the lack of fine-tuning.

\smallskip

\textbf{LLMs agreement (RQ2).} 
Across all 30 scenarios, pairwise model agreement was 73.3\% for TCR and 86.7\% for BAR. Agreement varied across scenarios, with high divergence in ethically ambiguous situations especially those involving trade-offs between duties, rights, or risks.

\noindent
Figure~\ref{fig:sTCR_BAR_table} visualizes model agreement using interpretive thresholds inspired by Fleiss’ Kappa~\cite{falotico2015fleiss}. Green and yellow cells mark strong and fair agreement, while red signals interpretive divergence. This allows rapid identification of high-variance scenarios that may warrant escalation or human oversight. LLMs reach high consistency on moral acceptability and moderate-to-high consistency on ethical theory. Divergences occur in conceptually complex cases, confirming that model disagreements align with known areas of ethical ambiguity.

\smallskip

\textbf{LLMs vs. Human Experts (RQ3).}
To compare human and model behavior, we compute the combined z-score per scenario for both groups. 
%required by major revision
\major{In several cases (e.g., scenarios numbered 14, 15, 21, 27, 29 in the replication package), LLM agreement patterns parallel the degree of expert convergence on the same scenarios, whereas sustained divergences on both sides (e.g., scenarios 4, 9, 11, 13, 17, 23) indicate intrinsic interpretive ambiguity. We therefore interpret agreement as a signal of stability, and disagreement as a cue for triage, rather than as evidence of normative correctness. Expert TCR values (often around two-thirds) should not be read as poor performance but as human-level pluralism under minimal normative instruction. This variability is a feature of the task design: it exposes multiple legitimate framings and makes explicit where automated interpretation should escalate to humans.}
%In many cases, high agreement among LLMs mirrors high agreement among experts (e.g., scenarios numbered 14, 15, 21, 27, 29 in the replication package). Shared low scores (e.g., scenarios 4, 9, 11, 13, 17, 23) suggest persistent interpretive uncertainty. LLMs show comparable agreement to human experts across most scenarios. Convergences and divergences between groups align with scenario difficulty, indicating that model disagreement often reflects intrinsic ambiguity rather than random noise. Some scenarios, like scenario 3 %\mashal{maybe we refer to where this scenario 3 can be found, if you refer to the three bullets we have above, then we need to number the bullets, maybe?}, show higher model alignment than human consensus, and vice versa. These mismatches indicate both the strength and blind spots of LLM-based reasoning, and support human-in-the-loop configurations.

\smallskip

\textit{Single LLM vs. Ensemble.}
No single model was in perfect agreement with the modal ensemble across all scenarios. Several models exhibited idiosyncratic choices, suggesting that relying on a single LLM for ethically-sensitive decisions introduces variance and potential bias. Ensemble-based aggregation mitigates this by producing more robust and explainable outcomes, particularly in ethically ambiguous or high-stakes settings.

\smallskip

\textit{Implications for Software Engineering.}
The described capacity to identify scenarios with high or low model agreement is potentially actionable in SE pipelines. Scenarios with high agreement can be handled autonomously; those with low agreement can trigger alerts or escalate decisions for human review. Moreover, scenario-wise agreement scores can serve as confidence metrics to support ethical decision auditing, runtime triage in recommender or assistant systems, and automated filtering of morally unstable outputs. This sets the stage for embedding LLMs as modular ethical profilers providing scalable, explainable, and context-sensitive reasoning components in future software systems.

\smallskip

\begin{comment}

%we have a limitations paragraph in the discussion section

\textit{Limitations.}
Despite promising results, some limitations constrain this analysis: only three ethical theories are considered, excluding other frameworks (e.g., care ethics, relativism); models are prompted independently in zero-shot; no interaction, clarification, or user context is allowed; scenario phrasing and prompt structure may influence model behavior; prompt sensitivity was not systematically assessed; agreement metrics capture convergence, not correctness; systematic errors may go undetected. These limitations motivate not only the next section but also future work on theory extension, scenario design, and semantically robust evaluation tools.
\end{comment}
\section{Qualitative Analysis of LLM Explanations}
\label{sec:qualitative}

%\subsection{Motivation and Analytical Strategy}

The ability of an LLM to select an appropriate ethical theory or to judge an action's acceptability (as quantitatively analyzed in Section~\ref{sec:quantitative}) is only part of what constitutes ethical reasoning. In software engineering contexts where systems must explain decisions to users, auditors, or regulators, the \textit{explanatory layer} becomes essential. That is, this section addresses \textbf{RQ4} by complementing our quantitative agreement analysis with a qualitative investigation of the linguistic and conceptual properties of the explanations generated by LLMs. The structure, content, and coherence of such explanations directly affect the trustworthiness, transparency, and usefulness of the AI system involved. We aim to determine whether models produce morally meaningful, internally coherent, and theory-consistent explanations. We also explore how such explanations vary across models, and whether surface-level linguistic diversity masks deeper conceptual alignment. To this end, we design a multi-tiered qualitative analysis based on investigations that include %lexically and syntactically
\major{lexical and syntactic} diversification of the explanations, low lexical similarity implications for conceptual divergence, consistency of explanations with the ethical theory selected by the model, moral vocabularies and normative traditions emerging across the corpus. To answer these, we combine computational techniques (TF-IDF similarity, LDA topic modeling, clustering, dimensionality reduction) with structured manual review. This dual approach balances scale and interpretability, enabling us to identify both aggregate trends and fine-grained misalignments.

\smallskip

\begin{figure*}[t]
\centering
\begin{minipage}{0.6\textwidth}
  \centering
  \includegraphics[width=\linewidth]{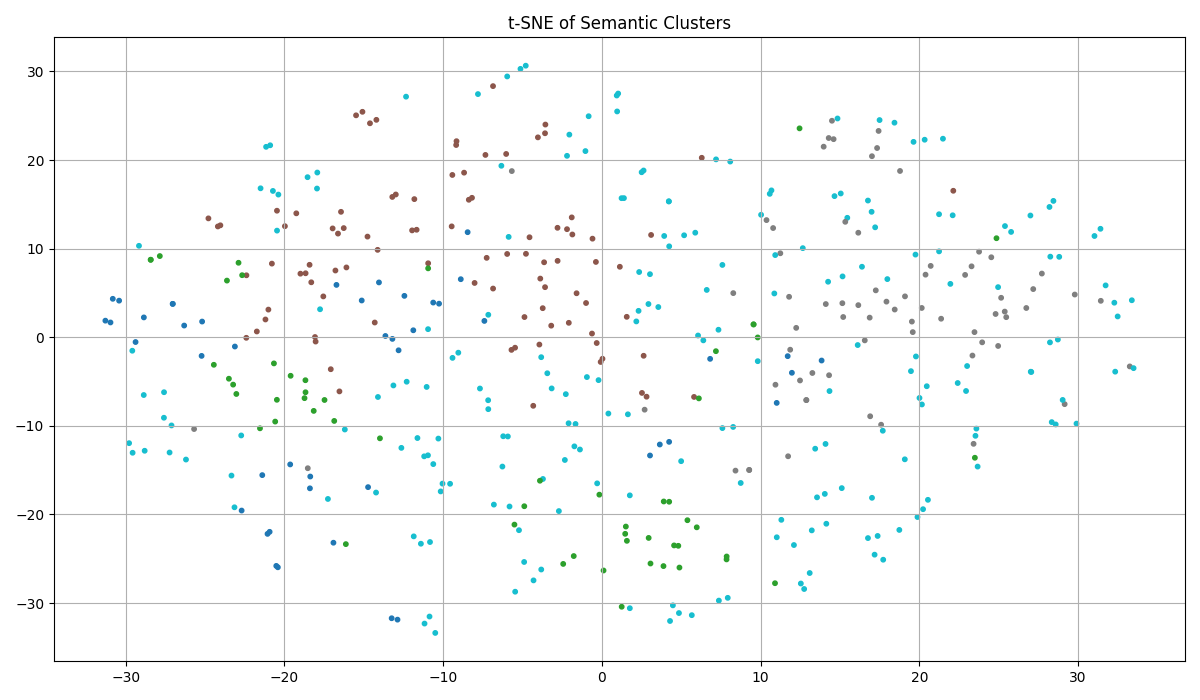}
  \caption{t-SNE Clustering}
  \label{fig:tsneclustering}
\end{minipage}%
\hfill
\begin{minipage}{0.34\textwidth}
  \centering
  \includegraphics[width=\linewidth]{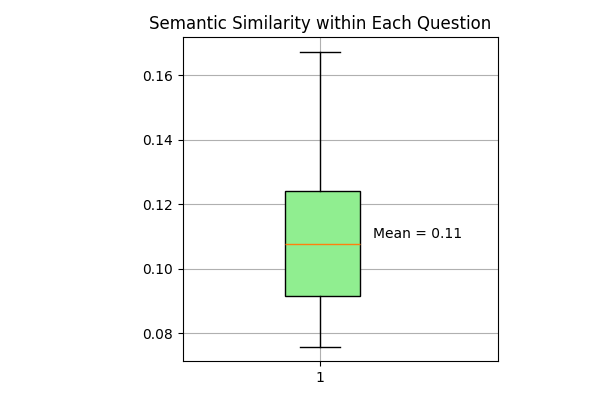}
  \vspace{-0.6cm}
  \caption{Similarity per question}
  \label{fig:similarityQ}
\end{minipage}
\vspace{0.5cm}
\end{figure*}

\textit{Lexical Diversity and Similarity.} We first assess lexical similarity using TF-IDF vectorization and pairwise cosine similarity, computed both across all explanations globally and within each scenario. Despite a consistent prompt structure and fixed task format, we observe substantial variation in surface form. The average pairwise cosine similarity across scenarios is 0.11, with a min of 0.02 and a max of 0.17 (Figure~\ref{fig:similarityQ}). These low values indicate that models rarely repeat phrasings, even when agreeing on the same theory and judgment. This suggests that explanations are 
%not rigidly templated or memorized, but are 
generated with contextual variation. This result implies that LLMs are not simply giving fixed moral explanations but are capable of producing distinct, scenario-sensitive moral language. It also poses a methodological challenge; similarity metrics that rely on surface overlap will systematically underestimate conceptual agreement unless paraphrase-aware techniques are employed.

%\smallskip

\textit{Semantic Clustering and Model Positioning.} To understand how explanations vary semantically, we applied Principal Component Analysis (PCA) and t-distributed Stochastic Neighbor Embedding (t-SNE) to the TF-IDF vector space. These projections allow us to visualize clusters of models and identify potential outliers.
\begin{figure}[ht]
\centering
\includegraphics[width=0.5\textwidth]{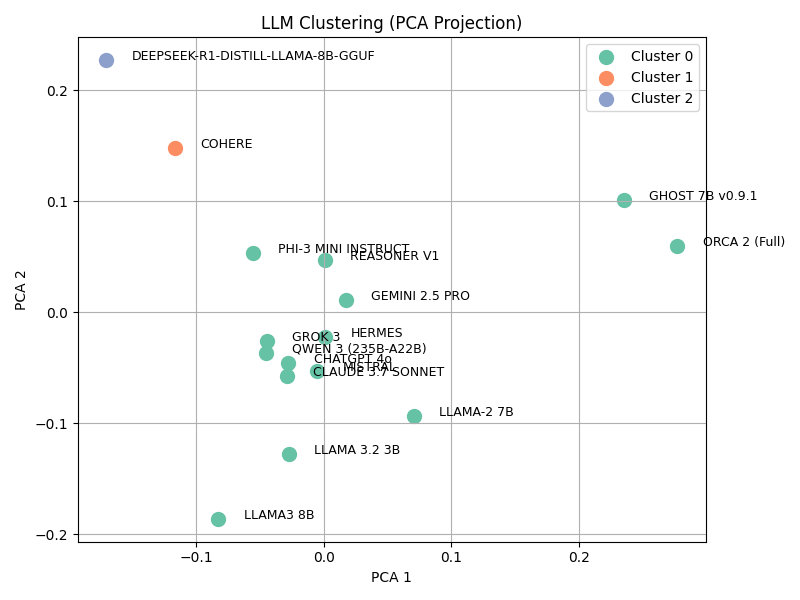}
\vspace{-0.6cm}
\caption{PCA Clustering}
\label{fig:pcaclustering}
\end{figure}
As shown in Figures~\ref{fig:pcaclustering} and~\ref{fig:tsneclustering}, most models occupy a dense central region, indicating that their explanations are semantically similar at the coarse level. A few models, notably COHERE and DEEPSEEK-R1, appear in peripheral regions. Manual inspection reveals that this divergence is primarily due to stylistic verbosity or recurrent syntactic structures, rather than shifts in ethical stance. We find no evidence that any model persistently aligns with a particular ethical theory in its language use alone. Instead, model positioning appears to reflect stylistic preferences rather than moral commitments. This further supports the hypothesis that surface diversity does not equate to conceptual inconsistency.
\begin{figure}[ht]
\centering
%\includegraphics[width=0.5\textwidth]{figures/qualitative/plot_llm_heatmap2.png}
%
%
%
% cropped out the legend to increase the labels font as requested by RevB, the numbers are in the boxes so the legend is not needed
%
%
%
\includegraphics[width=0.50\textwidth, clip, trim=0 0 88 0]{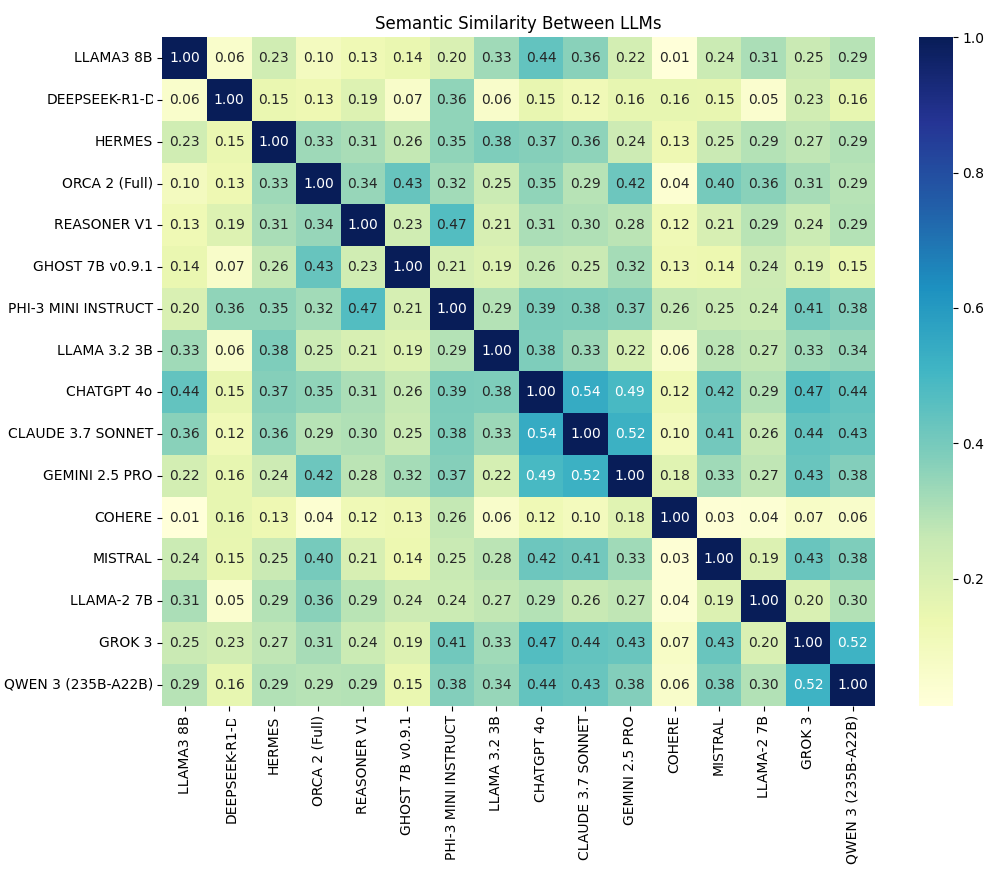}
\caption{Semantic similarity heatmap}
\label{fig:heatmap}
\end{figure}
The heatmap in Figure~\ref{fig:heatmap} confirms that there is no large cluster of highly similar models; most LLMs are only weakly related to each other in their use of language, specifically LLMs with the same family derivation.

\begin{figure}[ht]
\centering
\includegraphics[width=0.5\textwidth]{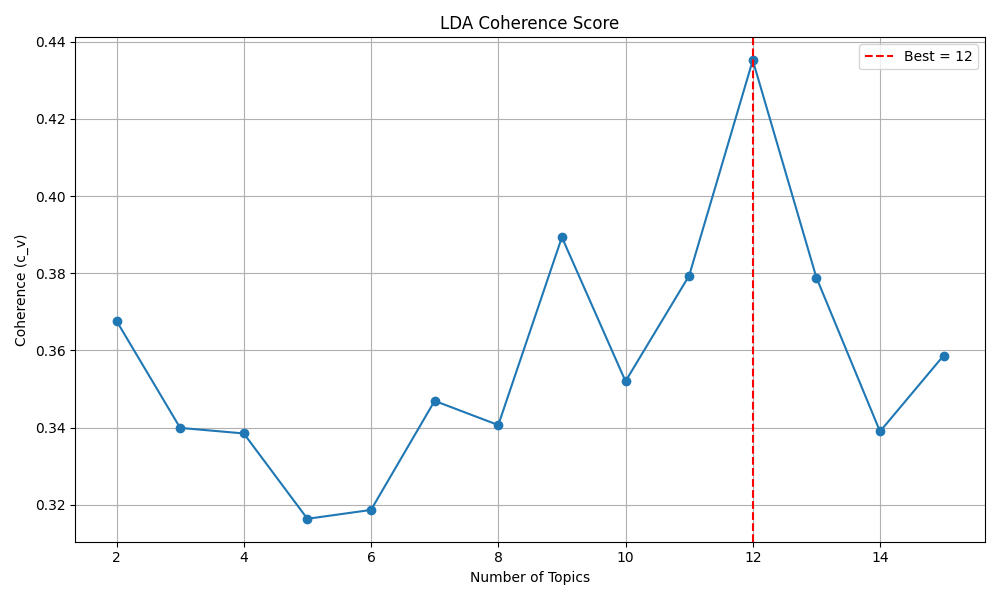}
\caption{LDA Coherence Score}
\label{fig:ldacoherence}
\end{figure}

\begin{figure}
\centering
\includegraphics[width=0.5\textwidth]{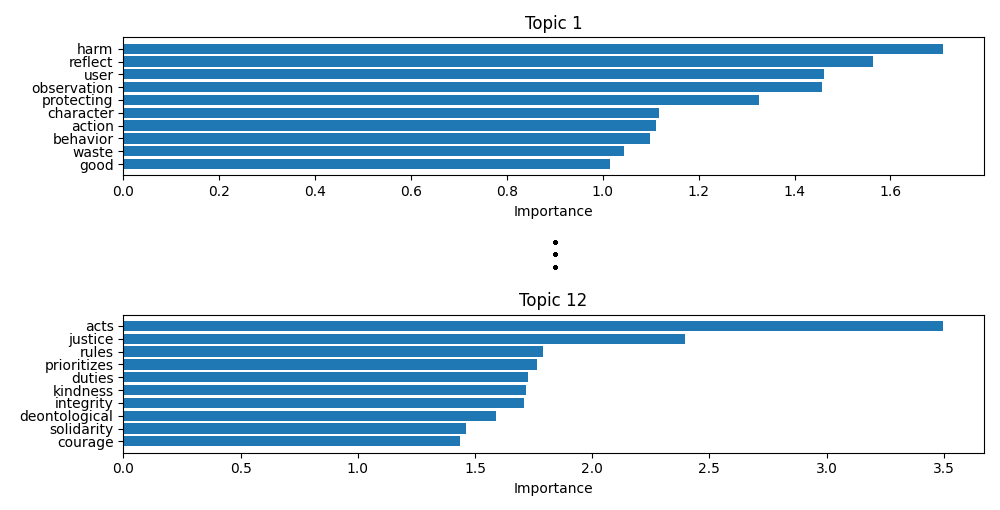}
\caption{\major{LDA topic excerpt. The full image is available in the replication package.}}
\vspace{0.5pt}
\label{fig:ldatopics}
\end{figure}

\smallskip
\major{
\textit{Topic Modeling and Moral Vocabulary.} We applied Latent Dirichlet Allocation (LDA) to the corpus of model explanations to uncover recurrent moral themes and to support downstream, user-directed interpretation. We used a standard preprocessing pipeline (lower-casing, tokenization, stopword removal, lemmatization, and bigram detection), built a filtered dictionary and bag-of-words corpus, and trained LDA models for a range of topic counts \(k\). Model selection was guided by the coherence curve \(C_v\), which improved up to \(k=12\) and then showed negligible gains, indicating that twelve topics provide an adequate balance between semantic granularity and interpretability (Figure~\ref{fig:ldacoherence}). This choice is therefore empirically grounded rather than arbitrary. For each topic we inspected the highest-weight tokens and representative explanations nearest to the topic centroid to propose a short, human-readable descriptor. Crucially, this labeling is descriptive and domain/application-oriented: it illustrates how an analyst or practitioner can use our system to surface moral vocabulary and attach domain-specific meanings, rather than asserting a fixed ontology. We report some of the labeling applied to the actual clusters (Figure \ref{fig:ldatopics}), the complete labeling table, including the token-to-label mapping and per-topic exemplar explanations, is provided in the replication package.}
\begin{itemize}
    \item \major{1 - Harm and character responsibility (Tokens such as \textit{harm, protecting, character, behavior, good} indicate emphasis on avoiding harm through responsible conduct.)}
    \item \major{2 - Normative framing and intervention (Presence of \textit{deontological, virtue, privacy, safety, intervening} reflects meta-theoretical framing and protective action.)}
    \item \major{...}
    \item \major{12 - Justice, integrity, and solidarity (\textit{justice, rules, duties, integrity, solidarity, courage} signals rule-guided fairness and character strength.)}
\end{itemize}
\major{\noindent Importantly, no single topic dominates the corpus. Explanations often blend multiple topics within a single rationale, indicating flexible use of moral vocabulary across theories. This pluralism is expected in ethical discourse and is compatible with our goal: the system exposes structured moral themes and their lexical supports, while leaving the interpretive labeling to the analyst’s aims and domain constraints.}
%\textit{Topic Modeling and Moral Vocabulary.} To ensure that the topic modeling reflected meaningful and stable moral themes, we evaluated LDA coherence across varying numbers of topics. Coherence scores quantify the degree to which the most probable words within each topic co-occur in the corpus, and thus provide a proxy for semantic interpretability. As shown in Figure~\ref{fig:ldacoherence}, coherence increased monotonically up to twelve topics and then plateaued, suggesting that twelve topics capture the major explanatory structures without overfitting or fragmenting the semantic space.
%The selection of twelve topics is therefore grounded in empirical optimization, rather than arbitrary selection, and allows us to balance granularity with interpretability. Each topic, when qualitatively examined, corresponds to recognizable ethical constructs (e.g., obligation, utility, virtue, autonomy), reinforcing the validity of the LDA decomposition. This ensures that subsequent interpretation and alignment with ethical traditions are based on a semantically coherent foundation.
%To explore deeper conceptual content, we applied Latent Dirichlet Allocation (LDA) to the corpus of explanations, aiming to uncover latent moral themes and trace their relation to ethical traditions. The model was optimized for coherence, which stabilized at twelve topics. The resulting topics (Figure~\ref{fig:ldatopics}) correspond to recognizable moral constructs: duty, virtue, honesty, justice, social utility, obligation, empathy, and harm minimization. Importantly, no single topic dominates the corpus. Most explanations blend multiple topics within a single explanation, indicating that models draw flexibly from different ethical vocabularies depending on the scenario. For example, a model might cite “public interest” (utilitarian), “obligation to disclose” (deontological), and “honesty” (virtue ethics) in the same explanation. %This pluralism mirrors real-world moral reasoning and suggests that models are not rigidly mapping to predefined theory templates, but instead construct morally plausible arguments using a hybrid vocabulary.
%This pluralism reflects real-world moral reasoning and suggests that the models are not rigidly following predefined theoretical frameworks, but instead construct morally plausible arguments using a hybrid moral vocabulary.

\smallskip

\textit{Theory–Explanation Alignment.} A core requirement for ethical reasoning is that the explanation should be consistent with the moral theory being invoked. To assess this, we conducted a manual alignment check on a stratified sample of 180 responses (6 models × 10 scenarios × 3 ethical theories), manually excluding outliers. In over 90\% of cases, the explanation supported the selected theory in a coherent manner. For example, Deontological responses often cited duties, rules, or rights (e.g., “The action violates the duty of confidentiality.”), Utilitarian responses referenced consequences and well-being (e.g., “The outcome benefits more people.”), and Virtue-based explanations appealed to character or intention (e.g., “It reflects compassion and honesty.”). Misalignments, when \major{they} occurred, tended to arise in edge cases or procedurally complex scenarios. Occasionally, a model labeled a decision as “virtue ethics” but explained it in outcome-based terms. Mismatches were rare and not concentrated in specific models.

\smallskip

\begin{figure}[ht]
\centering
\includegraphics[width=0.53\textwidth]{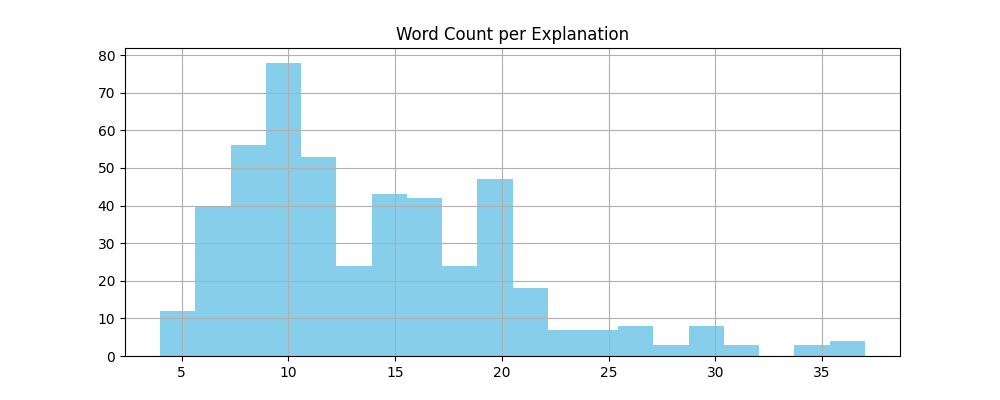}
\vspace{-0.6cm}
\caption{\major{Word} counts}
\label{fig:wordscounts}
\end{figure}

\textit{Conciseness and Structural Patterns.}
\begin{figure}[ht]
\centering
\includegraphics[width=0.53\textwidth]{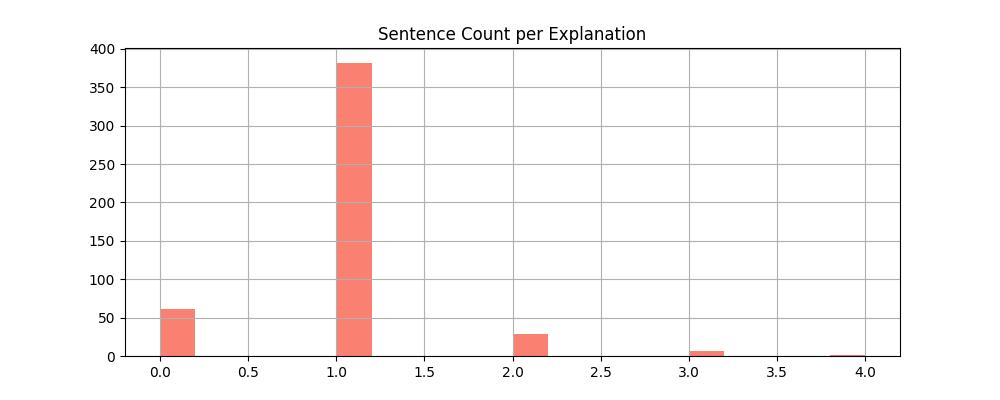}
\vspace{-0.6cm}
\caption{Sentence counts}
\label{fig:sentencecounts}
\end{figure}
We analyzed sentence and word counts across all explanations to understand the structural economy of model explanations. Over 95\% of outputs were a single sentence. Mean word counts ranged from 7.7 to 21.2 tokens (Figures~\ref{fig:sentencecounts}–\ref{fig:wordscounts}). Some models (e.g., GHOST, DEEPSEEK) tended to be more verbose, but longer explanations did not correlate with stronger theory alignment or clarity.
This brevity is notable; despite being concise, most explanations successfully reference relevant moral principles. The findings suggest that LLMs are able to generate ethical explanations that are both succinct and substantively meaningful\major{,} a valuable property for deployment in constrained SE contexts, such as UI prompts or trace logs.

\smallskip

\textit{Explanatory Convergence and Disagreement.}
\begin{figure}[ht]
\hspace*{-1.2em}
\includegraphics[width=0.525\textwidth]{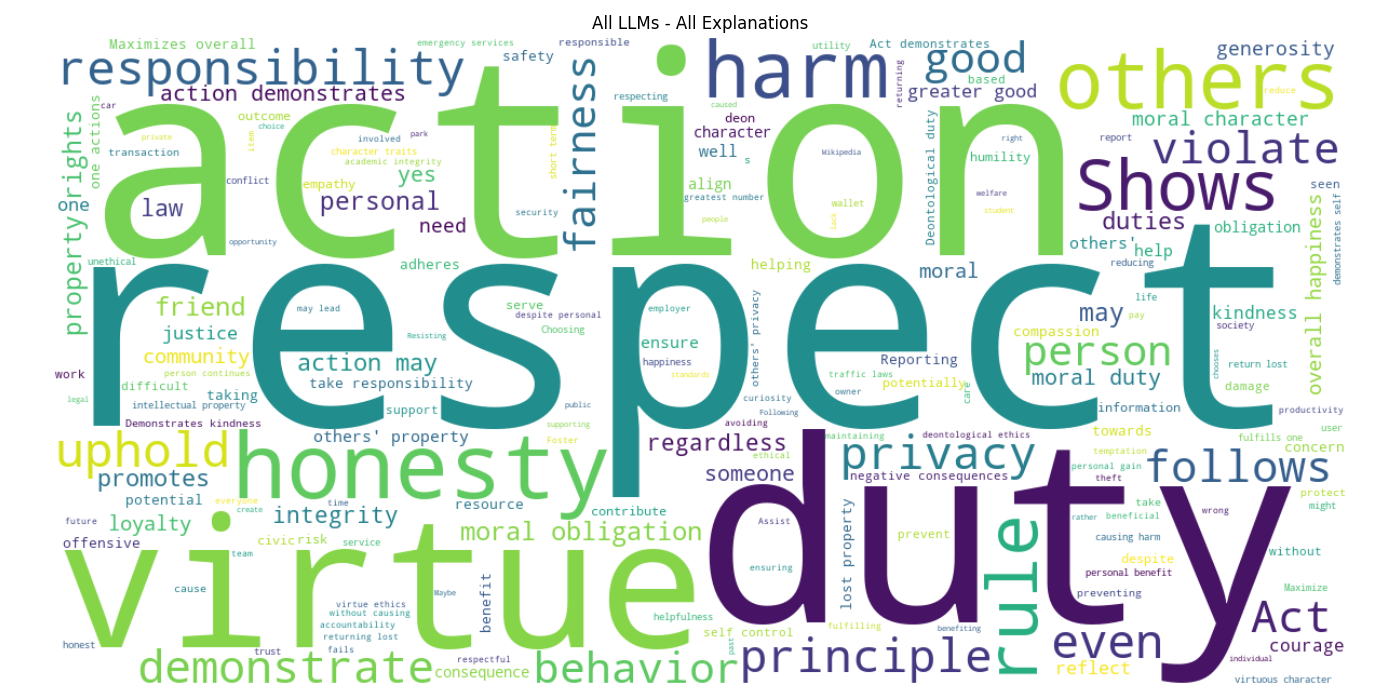}
\caption{Word cloud}
\label{fig:wordcloud}
\end{figure}
In high-consensus scenarios (e.g., helping a clerk, donating to Wikipedia), all models agreed both on judgment and theory, and their explanations while lexically distinct shared moral themes like civic virtue, altruism, or public benefit. Word cloud analysis (Figure~\ref{fig:wordcloud}) confirms recurring terms such as “respect”, “action”, “duty”, “virtue”, and “harm”. In low-agreement scenarios, such as those involving conflicting obligations or ambiguous responsibility, we observed both linguistic and conceptual spread. Some models emphasized personal integrity, others systemic outcomes. However, even here, the diversity appeared to reflect the ambiguity of the scenario rather than noise or error.

\smallskip

\textit{Implications for Ethical Profiling and SE Systems.}
These findings have direct implications for the use of LLMs in SE applications that require ethical reasoning or user profiling:

\begin{itemize}[left=0pt]
    \item \textit{Trust and Transparency.} The presence of coherent, interpretable explanations supports use cases requiring traceable decision-making (e.g., user-facing ethical explanations or compliance logging).
    \item \textit{Profile Construction.} The diversity and flexibility in moral vocabulary provide rich semantic data for generating dynamic ethical profiles based on user input.
    \item \textit{Model Assessment.} Surface similarity metrics may underestimate performance; systems should instead incorporate theory-alignment checks or conceptual similarity measures (e.g., based on semantic entailment).
    \item \textit{Ethical Memory.} Explanatory patterns may be used to build cumulative ethical profiles over time, enabling systems to adapt to evolving user values without retraining.
\end{itemize}

\textbf{Qualitative characteristics (RQ4).} From what emerges from our analysis, the LLM-generated explanations for ethical decisions are marked by high surface diversity and consistent moral coherence. Despite low lexical overlap across outputs, with average pairwise similarity scores around 0.11, the models consistently construct explanations that align with the selected ethical theory in over 90\% of sampled cases. %This indicates that models do not rely on fixed templates but instead generate context-sensitive explanations. 
In line with what has been argued so far, this also suggests that models operate beyond rigid theoretical frameworks, producing context-sensitive explanations.
Dimensionality reduction and clustering analyses (PCA, t-SNE) show that most models occupy a dense central region in semantic space, suggesting convergence at the conceptual level, even as their syntactic forms diverge. Outliers (e.g., COHERE, DEEPSEEK) diverge primarily due to stylistic rather than ethical differences. Topic modeling reveals that LLMs implicitly draw from a pluralistic moral vocabulary, spanning utilitarian, deontological, and virtue-based notions. Most explanations combine references to multiple moral principles within a single response, confirming the models' ability to construct hybrid, plausible explanations. %rather than mapping rigidly to single-theory templates. 
Structurally, over 95\% of explanations are a single sentence, and brevity does not preclude moral relevance. Even short explanations tend to highlight key ethical features (e.g., duty, harm, empathy), making them well-suited for constrained SE contexts. Finally, in high-consensus scenarios, lexical variation coexists with thematic convergence around shared moral terms (e.g., ``altruism'', ``civic duty'', ``harm''), while disagreement in complex cases reflects underlying ambiguity rather than interpretive noise. These findings support the use of LLMs for both traceable decision-making and dynamic ethical profiling in software engineering systems.

\section{Discussion}
\label{sec:discuss}

\textbf{RQ1.} Our findings provide evidence that state-of-the-art LLMs can engage in ethical reasoning when presented with complex, real-world made \major{explanations of acceptability}. Without fine-tuning or examples, models consistently identified the most applicable ethical theory and made acceptability explanations with substantial inter-model agreement. This capability suggests that LLMs possess an implicit grasp of moral reasoning principles, grounded in their pretraining on large-scale textual corpora. From an SE perspective, this opens the door to using LLMs as ethical reasoning modules in decision-making pipelines, such as requirement negotiation, user modeling, or system auditing.

%\subsection{RQ2: Do LLMs Reason Consistently Across Models and Scenarios?}

\smallskip

\textbf{RQ2.} Quantitative results showed that models converge more strongly on binary moral acceptability (86.7\% BAR) than on ethical theory classification (73.3\% TCR). While this difference reflects the higher abstraction level of theoretical judgments, the level of agreement observed is non-trivial. The scenario-dependent variability in TCR reveals an important feature: model disagreement tends to reflect ethical ambiguity inherent in the scenario rather than arbitrary noise. This suggests that ensemble disagreement can be used as a proxy for moral uncertainty, enabling software systems to trigger escalation or human intervention when LLMs disagree sharply.

%\subsection{RQ3: How Do LLMs Compare to Human Experts?}

\smallskip

\textbf{RQ3.} %Our comparison revealed substantial alignment between LLMs and human experts. 
\major{Overall, LLMs exhibit non-trivial agreement with experts that is more pronounced in prevalent classes and weaker on rare or edge cases.} Scenarios that elicited strong agreement among experts tended to also show high inter-model LLM agreement, and vice versa. This convergence reinforces the reliability of LLMs in interpreting familiar or structurally simple moral scenarios. Divergences, especially in edge cases, underscore the importance of hybrid systems that combine automated reasoning with human oversight. For SE applications involving legal, regulatory, or safety-critical implications, LLM-based profiling should not be deployed as an isolated decision-maker but as a complementary module.

%\subsection{RQ4: What Characterizes the Structure of LLM Moral Expanations?}

\smallskip

\textbf{RQ4.} Qualitative analyses demonstrated that LLMs generate explanations that are lexically diverse but conceptually coherent. Despite low textual similarity across models, explanations consistently aligned with the chosen moral theory in over 90\% of cases. Models blended terminology from multiple ethical traditions in natural, context-sensitive ways, %mirroring 
\major{tending to reflect} how human reasoners combine principles, consequences, and character-based considerations. This expressive flexibility is critical for ethical profiling, as it enables the detection of user-aligned reasoning patterns across different moral framings. Moreover, the compactness of most explanations (single sentences) and their theoretical consistency suggest that LLMs are capable of producing tractable, auditable moral outputs suitable for runtime interpretation and logging.
%required by metareview
\begin{tcolorbox}[colback=gray!10,
                  colframe=black,
                  arc=4mm,
                  boxrule=0.8pt,
                  left=2mm, right=2mm, top=1mm, bottom=1mm]
                  \major{\textbf{Agreement $\ne$ correctness.} Our design surfaces stability and ambiguity signals, it does not certify normative accuracy. In SE practice, high agreement supports automation with audit, while low agreement recommends human-in-the-loop escalation.}
\end{tcolorbox}                  

\smallskip

\textbf{Limitations and Scope of Validity.} While promising, our findings are bounded by some limitations:

\smallskip

\noindent\textit{Theoretical coverage.} We focus on three major ethical theories utilitarianism, deontology, and virtue ethics due to their widespread adoption in software engineering practice and education~\cite{vaniea2018securitytrolley}. These ethical theories provide well-established foundations for analyzing ethical dilemmas in technology contexts. While alternative theories such as care ethics or contractualism are less commonly applied, they offer valuable perspectives that could enrich ethical analyses. Future work may explore the integration of these additional frameworks to capture a broader spectrum of moral reasoning in software engineering.

\smallskip

\noindent\textit{Scenario framing.} Our prompts use concise, decontextualized scenarios. Richer formats (e.g., dialogues, system logs) may affect model interpretation. Our current prompts are decontextualized statements. An important next step is to apply the same ethical reasoning pipeline (Figure~\ref{fig:approach2}) to richer input modalities, including:
(i) chat transcripts from developer-agent interactions;
(ii) logs of user decisions in ethically sensitive configurations;
(iii) behavioral signals from simulation environments or system telemetry.
This would move the profiling process closer to real-time, context-aware ethical inference.

\smallskip

\noindent\textit{Zero-shot constraints.} All reasoning is performed without memory or clarification. Interactive or multi-turn reasoning may yield different profiles. An ethical profile need not be static. As users interact with a system, their decisions may reveal shifts in priorities, trade-offs, or ethical boundaries. Future work should implement an ethical memory module that incrementally updates a user's profile over time, capturing both stable dispositions and contextual shifts. This requires designing a temporal profiling architecture that tracks ethical indicators across scenarios and resolutions.

\smallskip
   
\noindent\textit{Agreement $\ne$ correctness.} Convergence does not imply normative accuracy. Human biases and model alignment may coincide but remain ethically questionable. In real deployments, users may reject or revise the moral judgments made by the system. Building on our current architecture, we envision an interactive loop in which: (i) the system proposes an ethical explanation; (ii) the user confirms, modifies, or rejects the reasoning; (iii) the profile is updated accordingly. This would enable both user agency and model refinement over time, reducing the risk of misaligned ethical personalization. 

\smallskip

\major{\noindent\textit{Prompt sensitivity.} Our zero-shot, single-turn protocol deliberately controls for instruction complexity; however, model behavior can still be sensitive to seemingly innocuous variations in prompt phrasing, formatting, or input length. We therefore treat prompt sensitivity as a threat to validity and an explicit boundary of our claims. A systematic sensitivity analysis is left as future work. In practice, we recommend freezing prompt templates in repositories and reporting all formatting details that might affect reproducibility.}

\smallskip

Even if the findings support the potential viability, these limitations suggest caution in direct deployment and highlight the need for further validation before integrating LLM-based profiling into high-stakes SE systems. Our results position LLMs as viable components for modular ethical reasoning in SE. Possible use cases include: \textit{decision auditing} for moral rationales generation for SE tool outputs (e.g., in requirements prioritization or resource allocation); \textit{autonomy triage} to route decisions to humans when LLMs disagree, reducing risk in ethically charged contexts; \textit{agent personalization} to tailor behavior of autonomous SE agents based on learned ethical user profiles. More broadly, the ability to extract consistent moral structure from language enables a shift from static ethics-as-checklists to adaptive, traceable, and user-aligned ethical cognition in engineered systems.
\section{Related Work}
\label{sec:related}
\noindent\emph{\major{Ethics in autonomous and software-intensive systems.}}
The integration of ethical considerations into autonomous systems has been widely examined in software engineering. Prior work spans design-time approaches that encode ethics via codes of conduct, principles, and rules~\cite{suri2023software,alidoosti2021ethics,alidoosti2022incorporating,DBLP:journals/access/ShahinHNPSGW22,trailer2024ciniselli}, and verification-time approaches that formalize and check system decisions against ethical frameworks~\cite{jedlickova2024ensuring,dennis2016formal,cardoso2021implementing,inverardi2019ethics,de2024engineering,inverardi2022ethical,Machine_Ethics_in_Changing_Contexts:2021,karim2017ethical}. This body of research establishes both the need and the mechanisms for ensuring ethically compliant behavior in autonomous decision pipelines.

%\smallskip
\noindent\emph{\major{LLMs in software engineering and the need for ethical reasoning.}}
Concurrently, LLMs have been adopted across SE tasks such as code generation, bug detection, and documentation~\cite{hou2024large}. As these models are integrated into SE workflows, it becomes important to assess whether they exhibit ethical reasoning and how they might be leveraged in systems with ethical implications~\cite{DBLP:conf/emnlp/RaoKTAC23,han2022aligning}.

%\smallskip
\noindent\emph{\major{Model-centric evaluations of moral reasoning.}}
Han et al.~\cite{han2022aligning} evaluate LLM understanding of moral/ethical reasoning across five domains (justice, deontology, virtue ethics, utilitarianism, and commonsense morality). They fine-tune BERT-base, BERT-large, RoBERTa-large, and ALBERT-xxlarge on ETHICS datasets comprising over 13{,}000 scenarios, and then test the models’ ability to classify scenarios in line with the ethical theories used during fine-tuning. In contrast, the present work does not rely on fine-tuning; it evaluates whether pre-trained LLMs, in a zero-shot setup, display ethical reasoning.

%\smallskip
\noindent\emph{\major{Value identification in scholarly texts vs. ethically charged situations.}}
A complementary thread leverages LLMs to extract values and structure from scientific writing. For example,~\cite{ICSE2025PAPER} studies ChatGPT’s ability to identify human values from titles and abstracts of SE publications using Schwartz’s theory~\cite{schwartz2012overview}, followed by manual verification by humans. Related efforts show that LLMs can assist in extracting insights from scholarly texts, classifying publications, and mining metadata for literature reviews~\cite{hou2024large,alshami2023harnessing}. These studies focus on value detection in academic prose rather than on evaluating ethical concrete reasoning, ethically charged scenarios.

%\smallskip
\noindent\emph{\major{Reasoning with policies and learning moral rewards.}}
Rao et al.~\cite{DBLP:conf/emnlp/RaoKTAC23} argue against hard-coding specific moral values in LLMs and advocate for general ethical reasoning capacities that adapt to diverse contexts. They introduce in-context ethical policies defined at varying abstraction levels and grounded in deontology, virtue ethics, and consequentialism, reporting experiments across GPT-3, ChatGPT, and GPT-4 with GPT-4 showing stronger ethical reasoning. Tennant et al.~\cite{DBLP:conf/iclr/TennantHM25} incorporate intrinsic moral rewards, grounded in deontological and utilitarian theories, into reinforcement learning fine-tuning. Using the Iterated Prisoner’s Dilemma, they demonstrate that LLM agents can learn morally aligned strategies and even unlearn previously selfish behaviors. While these works shape model behavior through policies or moral rewards, the present study compares outputs across multiple models without additional tuning, treating ethical reasoning as a testing ground rather than a direct optimization target.

%\smallskip
\noindent\emph{\major{Positioning of the present study.}}
Prior research establishes design/verification-time mechanisms for ethics in autonomous systems, documents LLM utility in SE, and explores both fine-tuned moral reasoning and in-context ethical policy use. The contribution here is orthogonal: a zero-shot assessment of pre-trained LLMs’ ethical reasoning on ethically charged scenarios, contrasting with fine-tuned or policy-conditioned settings, and distinct from value-mining in academic text.

\section{Conclusion and Future Work}
\label{sec:conclude}

Our work investigates the potential of leveraging LLMs as an ethical component within the software engineering pipeline to automate the generation of user ethical profiles. The profiles represent users' ethical preferences in a structured way, guiding system behavior to align its decisions with user values. To achieve this, we evaluated the ethical reasoning capabilities of 16 state-of-the-art Large Language Models (LLMs) by presenting them with 30 ethically charged scenarios. We prompted LLMs to identify which ethical theory among utilitarianism, deontology, and virtue ethics applies to the action detailed in the scenario, determine whether the action described is morally acceptable, and provide the reasoning behind their choice. We then computed inter-model agreement between the responses of the 16 LLMs using Theory Consistency Rate (TCR) and Binary Agreement Rate %metrics
(BAR) \major{metrics}. We replicated the same process with three expert ethicists and compared their responses with LLMs using z-score\major{s} to analyze the agreement between their responses. Moreover, we performed qualitative analysis of the LLM explanations to determine whether models produce meaningful and theory-consistent explanations. The results indicate LLMs' applicability within the software engineering pipeline to evaluate ethical contexts. Future work will prompt LLMs to identify actions based on a broader set of ethical theories, and will examine the applicability and interactions of these theories in more complex scenarios. Moreover, we plan to investigate how this framework could be integrated in existing SE toolchains.

\section*{Acknowledgments}
Authors wish to thank the three expert ethicists Prof. Simone Gozzano, Prof. Marco Segala, Prof. Massimiliano Palmiero and the entire Exosoul research group for their contributions to this paper.
This work has been partially funded by (a) the PRIN project P2022RSW5W - RoboChor: Robot Choreography, (b) the PRIN project 2022JKA4SL - HALO: etHical-aware AdjustabLe autOnomous systems, and (c) the PRIN project 2022JAFATE - CAVIA: enabling the Cloud-to-Autonomous-Vehicles continuum for future Industrial Applications; (d) PNRR MUR (Italy) -- Centro Nazionale HPC, Big Data e Quantum Computing, Spoke9 - Digital Society \& Smart Cities (grant CN\_00000013).
%\smallskip

%\vfill

\bibliographystyle{ieeetr}
\bibliography{ref}

\begin{thebibliography}{10}

\bibitem{suri2023software}
S.~Suri, S.~N. Das, K.~Singi, K.~Dey, V.~S. Sharma, and V.~Kaulgud, ``Software engineering using autonomous agents: Are we there yet?,'' in {\em 38th IEEE/ACM International Conference on Automated Software Engineering}, pp.~1855--1857, 2023.

\bibitem{jedlickova2024ensuring}
A.~Jedlickova, ``Ensuring ethical standards in the development of autonomous and intelligent systems,'' {\em IEEE Transactions on Artificial Intelligence}, vol.~5, no.~12, pp.~5863--5872, 2024.

\bibitem{InsightsISSRE}
P.~Pelliccione and N.~Laranjeiro, ``Insights from the software reliability research community,'' {\em IEEE Reliability Magazine}, vol.~1, no.~1, pp.~10--14, 2024.

\bibitem{waldman2019power}
A.~E. Waldman, ``Power, process, and automated decision-making,'' {\em Fordham L. Rev.}, vol.~88, p.~613, 2019.

\bibitem{sharma2022recent}
A.~Sharma, V.~Sharma, M.~Jaiswal, H.-C. Wang, D.~N.~K. Jayakody, C.~M.~W. Basnayaka, and A.~Muthanna, ``Recent trends in ai-based intelligent sensing,'' {\em Electronics}, vol.~11, no.~10, p.~1661, 2022.

\bibitem{anderson2018artificial}
J.~Anderson, L.~Rainie, and A.~Luchsinger, ``Artificial intelligence and the future of humans,'' {\em Pew Research Center}, vol.~10, no.~12, pp.~1--10, 2018.

\bibitem{autili2025engineering}
M.~Autili, M.~De~Sanctis, P.~Inverardi, and P.~Pelliccione, ``Engineering digital systems for humanity: a research roadmap,'' {\em ACM Transactions on Software Engineering and Methodology}, 2025.

\bibitem{de2024engineering}
M.~De~Sanctis and P.~Inverardi, ``Engineering ethical-aware collective adaptive systems,'' in {\em International Symposium on Leveraging Applications of Formal Methods}, pp.~238--252, Springer, 2024.

\bibitem{hendrycks2020aligning}
D.~Hendrycks, C.~Burns, S.~Basart, A.~Critch, J.~Li, D.~Song, and J.~Steinhardt, ``Aligning ai with shared human values,'' in {\em 9th International Conference on Learning Representations}, 2021.

\bibitem{alidoosti2022incorporating}
R.~Alidoosti, P.~Lago, E.~Poort, M.~Razavian, and A.~Tang, ``Incorporating ethical values into software architecture design practices,'' in {\em 19th International Conference on Software Architecture Companion}, pp.~124--127, 2022.

\bibitem{svegliato2021ethically}
J.~Svegliato, S.~B. Nashed, and S.~Zilberstein, ``Ethically compliant sequential decision making,'' in {\em AAAI Conference on Artificial Intelligence}, vol.~35, pp.~11657--11665, 2021.

\bibitem{tolmeijer2020implementations}
S.~Tolmeijer, M.~Kneer, C.~Sarasua, M.~Christen, and A.~Bernstein, ``Implementations in machine ethics: A survey,'' {\em ACM Computing Surveys}, vol.~53, no.~6, pp.~1--38, 2020.

\bibitem{inverardi2019ethics}
P.~Inverardi, ``Ethics and privacy in autonomous systems: A software exoskeleton to empower the user,'' in {\em Software Engineering for Resilient Systems: 11th International Workshop.}, pp.~3--8, 2019.

\bibitem{inverardi2022ethical}
P.~Inverardi, M.~Palmiero, P.~Pelliccione, and M.~Tivoli, ``Ethical-aware autonomous systems from a social psychological lens.,'' in {\em 6th International Workshop on Cultures of Participation in the Digital Age: {AI} for Humans or Humans for AI?}, pp.~43--48, 2022.

\bibitem{cervantes2020toward}
S.~Cervantes, S.~L{\'o}pez, and J.-A. Cervantes, ``Toward ethical cognitive architectures for the development of artificial moral agents,'' {\em Cognitive systems research}, vol.~64, pp.~117--125, 2020.

\bibitem{alidoosti2025exploring}
R.~Alidoosti, P.~Lago, M.~Razavian, and A.~Tang, ``Exploring the ethical landscape of software systems: A systematic literature review,'' {\em Journal of Systems and Software}, p.~112430, 2025.

\bibitem{jedlivckova2024ethical}
A.~Jedli{\v{c}}kov{\'a}, ``Ethical approaches in designing autonomous and intelligent systems: a comprehensive survey towards responsible development,'' {\em AI \& SOCIETY}, pp.~1--14, 2024.

\bibitem{bremner2019proactive}
P.~Bremner, L.~A. Dennis, M.~Fisher, and A.~F. Winfield, ``On proactive, transparent, and verifiable ethical reasoning for robots,'' {\em Proceedings of the {IEEE}}, vol.~107, no.~3, pp.~541--561, 2019.

\bibitem{winfield2014towards}
A.~F. Winfield, C.~Blum, and W.~Liu, ``Towards an ethical robot: internal models, consequences and ethical action selection,'' in {\em Conference towards autonomous robotic systems}, pp.~85--96, 2014.

\bibitem{Winfield2019}
A.~F.~T. Winfield and V.~V. Hafner, ``Anticipation in robotics,'' {\em Handbook of Anticipation: Theoretical and Applied Aspects of the Use of Future in Decision Making}, pp.~1587--1615, 2019.

\bibitem{alidoosti2021ethics}
R.~Alidoosti, ``Ethics-driven software architecture decision making,'' in {\em 18th International Conference on Software Architecture Companion}, pp.~90--91, 2021.

\bibitem{townsend2022pluralistic}
B.~Townsend, C.~Paterson, T.~Arvind, G.~Nemirovsky, R.~Calinescu, A.~Cavalcanti, I.~Habli, and A.~Thomas, ``From pluralistic normative principles to autonomous-agent rules,'' {\em Minds and Machines}, vol.~32, no.~4, pp.~683--715, 2022.

\bibitem{winter2019advancing}
E.~Winter, S.~Forshaw, L.~Hunt, and M.~A. Ferrario, ``Advancing the study of human values in software engineering,'' in {\em 2019 IEEE/ACM 12th International Workshop on Cooperative and Human Aspects of Software Engineering (CHASE)}, pp.~19--26, IEEE, 2019.

\bibitem{memon2023automated}
M.~A. Memon, G.~L. Scoccia, and M.~Autili, ``Automated negotiation-preliminary results of a systematic mapping study,'' in {\em 2023 38th IEEE/ACM International Conference on Automated Software Engineering Workshops (ASEW)}, pp.~94--99, IEEE, 2023.

\bibitem{MemonAFSI24}
M.~A. Memon, M.~Autili, G.~Filippone, G.~L. Scoccia, and P.~Inverardi, ``A high-level architecture of an automated context-aware ethics-based negotiation approach,'' in {\em Proceedings of the 39th {IEEE/ACM} International Conference on Automated Software Engineering, ({ASE})}, {ACM}, 2024.

\bibitem{donati2024representing}
D.~Donati, Z.~Assadi, S.~Gozzano, P.~Inverardi, and N.~Troquard, ``On representing humans' soft-ethics preferences as dispositions,'' in {\em Proceedings of the Ital-IA Intelligenza Artificiale}, vol.~3762 of {\em {CEUR} Workshop Proceedings}, pp.~135--140, 2024.

\bibitem{barn2016you}
B.~S. Barn, ``Do you own a volkswagen? values as non-functional requirements,'' in {\em Human-Centered and Error-Resilient Systems Development}, pp.~151--162, Springer, 2016.

\bibitem{alfieri2022exosoul}
C.~Alfieri, P.~Inverardi, P.~Migliarini, and M.~Palmiero, ``Exosoul: Ethical profiling in the digital world,'' in {\em HHAI2022: Augmenting Human Intellect}, pp.~128--142, IOS Press, 2022.

\bibitem{cacmInverardi2019}
P.~Inverardi, ``The european perspective on responsible computing,'' {\em Commun. ACM}, vol.~62, p.~64, mar 2019.

\bibitem{InverardiMP23}
P.~Inverardi, P.~Migliarini, and M.~Palmiero, ``Systematic review on privacy categorisation,'' {\em Computer Science Review}, vol.~49, p.~100574, 2023.

\bibitem{DiRuscioIMN24}
D.~Di~Ruscio, P.~Inverardi, P.~Migliarini, and P.~T. Nguyen, ``Leveraging privacy profiles to empower users in the digital society,'' {\em Automated Software Engineering}, vol.~31, no.~1, p.~16, 2024.

\bibitem{voigt2017eu}
P.~Voigt and A.~Von~dem Bussche, ``The eu general data protection regulation (gdpr),'' {\em A Practical Guide, 1st Ed., Cham: Springer International Publishing}, vol.~10, no.~3152676, pp.~10--5555, 2017.

\bibitem{EU_AI_Act_Proposal}
E.~Commission, ``Proposal for a regulation laying down harmonised rules on artificial intelligence (artificial intelligence act) and amending certain union legislative acts,'' 2021.

\bibitem{ai2019high}
H.~AI, ``High-level expert group on artificial intelligence,'' {\em Ethics guidelines for trustworthy AI}, vol.~6, 2019.

\bibitem{han2022aligning}
S.~Han, E.~Kelly, S.~Nikou, and E.-O. Svee, ``Aligning artificial intelligence with human values: reflections from a phenomenological perspective,'' {\em AI \& SOCIETY}, pp.~1--13, 2022.

\bibitem{alshami2023harnessing}
A.~Alshami, M.~Elsayed, E.~Ali, A.~E. Eltoukhy, and T.~Zayed, ``Harnessing the power of chatgpt for automating systematic review process: methodology, case study, limitations, and future directions,'' {\em Systems}, vol.~11, no.~7, p.~351, 2023.

\bibitem{DBLP:conf/emnlp/RaoKTAC23}
A.~Rao, A.~Khandelwal, K.~Tanmay, U.~Agarwal, and M.~Choudhury, ``Ethical reasoning over moral alignment: {A} case and framework for in-context ethical policies in llms,'' in {\em Findings of the Association for Computational Linguistics: {EMNLP} 2023, Singapore, December 6-10, 2023} (H.~Bouamor, J.~Pino, and K.~Bali, eds.), pp.~13370--13388, Association for Computational Linguistics, 2023.

\bibitem{DBLP:conf/iclr/TennantHM25}
E.~Tennant, S.~Hailes, and M.~Musolesi, ``Moral alignment for {LLM} agents,'' in {\em The Thirteenth International Conference on Learning Representations, {ICLR} 2025, Singapore, April 24-28, 2025}, OpenReview.net, 2025.

\bibitem{AutiliACCESS19}
M.~Autili, D.~Di~Ruscio, P.~Inverardi, P.~Pelliccione, and M.~Tivoli, ``A software exoskeleton to protect and support citizen's ethics and privacy in the digital world,'' {\em {IEEE} Access}, vol.~7, pp.~62011--62021, 2019.

\bibitem{ICSE2025PAPER}
D.~Mougouei, A.~Azarnik, M.~Fahmideh, E.~Mougouei, H.~K. Dam, A.~A. Khan, S.~Rafi, J.~A. Khan, and A.~Ahmad, ``A first look at ai trends in value-aligned software engineering publications: Human-llm insights,'' in {\em Proceedings of the 47th International Conference on Software Engineering: Software Engineering in Society, ICSE-SEIS2025}, pp.~82--93, {ACM}, 2025.

\bibitem{alfieri2023ethical}
C.~Alfieri, D.~Donati, S.~Gozzano, L.~Greco, and M.~Segala, ``Ethical preferences in the digital world: The exosoul questionnaire,'' in {\em HHAI2022: Augmenting Human Intellect}, pp.~290--299, IOS Press, 2023.

\bibitem{singh2023approaches}
S.~Singh, ``Approaches, theories, and role of ethics in computer science and engineering,'' in {\em 2023 Congress in Computer Science, Computer Engineering, \& Applied Computing (CSCE)}, pp.~01--08, IEEE, 2023.

\bibitem{UNESCOGuidelines}
{UNESCO}, ``Recommendation on the ethics of artificial intelligence,'' 2022.

\bibitem{ryan2020artificial}
M.~Ryan and B.~C. Stahl, ``Artificial intelligence ethics guidelines for developers and users: clarifying their content and normative implications,'' {\em Inf., Comm. and Ethics in Society}, vol.~19, no.~1, pp.~61--86, 2020.

\bibitem{guarini2013introduction}
M.~Guarini, ``Introduction: machine ethics and the ethics of building intelligent machines,'' {\em Topoi}, vol.~32, no.~2, pp.~213--215, 2013.

\bibitem{anderson2020machine}
M.~Anderson and S.~L. Anderson, ``Machine ethics: Creating an ethical intelligent agent,'' in {\em Machine ethics \& robot ethics}, pp.~237--248, Routledge, 2020.

\bibitem{o2012review}
M.~J. O’Fallon and K.~D. Butterfield, ``A review of the empirical ethical decision-making literature: 1996--2003,'' {\em Citation classics from the Journal of Business Ethics}, pp.~213--263, 2012.

\bibitem{mill2016utilitarianism}
J.~S. Mill, ``Utilitarianism,'' in {\em Seven masterpieces of philosophy}, pp.~329--375, Routledge, 2016.

\bibitem{hursthouse2017virtue}
R.~Hursthouse, ``On virtue ethics,'' in {\em Applied ethics}, pp.~29--35, Routledge, 2017.

\bibitem{sep-ethics-deontological}
L.~Alexander and M.~Moore, ``{Deontological Ethics},'' in {\em The {Stanford} Encyclopedia of Philosophy} (E.~N. Zalta, ed.), Metaphysics Research Lab, Stanford University, {W}inter 2021~ed., 2021.

\bibitem{boja2019user}
C.~Boja, A.~Zamfiroiu, M.~Zurini, and B.~Iancu, ``User behaviour profiling in social media applications,'' {\em Economic Computation \& Economic Cybernetics Studies \& Research}, vol.~53, no.~1, 2019.

\bibitem{gilbert2023rise}
J.~Gilbert, S.~Hamid, I.~A.~T. Hashem, N.~A. Ghani, and F.~F. Boluwatife, ``The rise of user profiling in social media: review, challenges and future direction,'' {\em Social Network Analysis and Mining}, vol.~13, no.~1, p.~137, 2023.

\bibitem{dong2021profiling}
X.~Dong, T.~Li, R.~Song, and Z.~Ding, ``Profiling users via their reviews: an extended systematic mapping study,'' {\em Software and Systems Modeling}, vol.~20, pp.~49--69, 2021.

\bibitem{falotico2015fleiss}
R.~Falotico and P.~Quatto, ``Fleiss’ kappa statistic without paradoxes,'' {\em Quality \& Quantity}, vol.~49, pp.~463--470, 2015.

\bibitem{vaniea2018securitytrolley}
K.~Vaniea and L.~J. Camp, ``Computer security trolley problems: Exploring key factors in security decision-making,'' in {\em Proceedings of the 2018 {IEEE} Symposium on Security and Privacy Workshops (SPW)}, pp.~134--140, IEEE, 2018.

\bibitem{DBLP:journals/access/ShahinHNPSGW22}
M.~Shahin, W.~Hussain, A.~Nurwidyantoro, H.~Perera, R.~Shams, J.~Grundy, and J.~Whittle, ``Operationalizing human values in software engineering: {A} survey,'' {\em {IEEE} Access}, vol.~10, pp.~75269--75295, 2022.

\bibitem{trailer2024ciniselli}
M.~Pezz\`{e}, M.~Ciniselli, L.~Di~Grazia, N.~Puccinelli, and K.~Qiu, ``The trailer of the {ACM} 2030 roadmap for software engineering,'' {\em SIGSOFT Softw. Eng. Notes}, vol.~49, p.~31–40, Oct. 2024.

\bibitem{dennis2016formal}
L.~Dennis, M.~Fisher, M.~Slavkovik, and M.~Webster, ``Formal verification of ethical choices in autonomous systems,'' {\em Robotics and Autonomous Systems}, vol.~77, pp.~1--14, 2016.

\bibitem{cardoso2021implementing}
R.~C. Cardoso, A.~Ferrando, L.~A. Dennis, and M.~Fisher, ``Implementing ethical governors in bdi,'' in {\em Workshop on Engineering Multi-Agent Systems}, pp.~22--41, 2021.

\bibitem{Machine_Ethics_in_Changing_Contexts:2021}
L.~A. Dennis, M.~M. Bentzen, F.~Lindner, and M.~Fisher, ``Verifiable machine ethics in changing contexts,'' in {\em AAAI Conference on Artificial Intelligence}, pp.~11470--11478, 2021.

\bibitem{karim2017ethical}
N.~S.~A. Karim, F.~Al~Ammar, and R.~Aziz, ``Ethical software: Integrating code of ethics into software development life cycle,'' in {\em 2017 International Conference on Computer and Applications (ICCA)}, pp.~290--298, IEEE, 2017.

\bibitem{hou2024large}
X.~Hou, Y.~Zhao, Y.~Liu, Z.~Yang, K.~Wang, L.~Li, X.~Luo, D.~Lo, J.~Grundy, and H.~Wang, ``Large language models for software engineering: A systematic literature review,'' {\em ACM Transactions on Software Engineering and Methodology}, vol.~33, no.~8, pp.~1--79, 2024.

\bibitem{schwartz2012overview}
S.~H. Schwartz, ``An overview of the schwartz theory of basic values,'' {\em Online readings in Psychology and Culture}, vol.~2, no.~1, pp.~10--20, 2012.

\end{thebibliography}

\end{document}